# Tensor-based reduction of linear parameter-varying state-space models

**Bogoljub Terzin**[1] | **E. Javier Olucha**[1] | **Amritam Das**[1] | **Siep Weiland**[1] | **Roland Tóth**[1,2]

[1]Control Systems Group, Eindhoven University of Technology, Eindhoven 5600MB, The Netherlands

[2]Systems and Control Lab, HUN-REN Institute for Computer Science and Control, 1111 Budapest, Hungary

**Correspondence**

Corresponding author: Bogoljub Terzin, Email: b.terzin@tue.nl

## Abstract

The Linear Parameter-Varying (LPV) framework is a powerful tool for controlling nonlinear and complex systems, but the conversion of nonlinear models into LPV forms often results in high-dimensional and overly conservative LPV models. To be able to apply control strategies, there is often a need for model reduction in order to reduce computational needs. This paper presents the first systematic approach for the joint reduction of state order and scheduling signal dimension of LPV state space models. The existing methods typically address these reductions separately. By formulating a tensorial form of LPV models with an affine dependency on the scheduling variables, we leverage tensor decomposition to find the dominant components of state and scheduling subspaces. We extend the common Petrov-Galerkin projection approach to LPV framework by adding a scheduling projection. This extension enables the joint reduction. To find suitable subspaces for the extended Petrov-Galerkin projection, we have developed two different methods: tensor-based LPV moment matching, and an approach through Proper Orthogonal Decomposition. Advantages of the proposed methods are demonstrated on two different series-interconnected mass-spring-damper systems with nonlinear springs: one primarily used for comparison with other methods and a more elaborate higher-order model designed to assess scalability.

## 1 | INTRODUCTION

Over the past few decades, significant advancements have been made in *nonlinear* (NL) and time-varying control theory, pushing the limits of engineering capabilities in accuracy, energy efficiency, and robustness. However, design principles based on *linear time-invariant* (LTI) systems have proven to be insufficient for achieving the increasingly demanding performance specifications. Interestingly, it has been observed that a set of linear approximations can be sufficient to describe the underlying dynamic behaviour of many nonlinear systems over an operating range. This observation, which emerged in the 1980s, led to the development of a modeling and control framework that aimed to extend powerful LTI methods to describe and regulate the behavior of nonlinear and time-varying dynamics of real-world systems. The resulting *linear parameter-varying* (LPV) framework, which initially was built around the concept of gain-scheduling [1], has become a well established approach with many succesfull applications from aerospace engineering till automotive systems.

Originally, gain-scheduling involved linearizing a nonlinear system at different operating points, resulting in a collection of local LTI models. For each of these local models, an LTI controller was designed and interpolated over the operating range. Although, this approach was widely adopted in industrial applications, it lacked guarantees of global stability for the overall controlled system [2]. Hence, to overcome these disadvantages, it was realized that many nonlinear systems could be directly converted into an LPV form without the need for linearization, through a concept called *LPV embedding* [3]. In the global LPV

**Abbreviations:** LPV, linear parameter-varying; SS, state-space. LTI, linear time-invariant. TV, time-varying. NL, nonlinear. FOM, full-order model. ROM, reduced-order model. SVD, singular value decomposition. POD, proper orthogonal decomposition embedding of nonlinear and time-varying systems, the system representation is reformulated such that the relationships between the inputs and outputs become linear, but dependent on a new signal $p$, which is a function of the state $x$ and the input $u$ via a scheduling map $\eta$, i.e., $p = \eta(x, u)$. Then, intentionally, this connection of $p$ to $x$ and $u$ via $\eta$ is disregarded, and it is assumed that variations of $p$ can occur independently within a set $\mathcal{P} \subseteq \eta(\mathcal{X}, \mathcal{U})$. While this results in linearity of the resulting relation, allowing the application of powerful analysis and control techniques, it comes at the cost of introducing conservatism in the embedding of the nonlinear behaviour. Specifically, there are solution trajectories of the new LPV description that do not exist in the original nonlinear system, and these must be addressed by the controller to ensure stability and the desired performance during design.

Control synthesis and analysis of LPV models, as well as LPV system identification, have become well-developed in the last few decades. However, often the models obtained with LPV embedding approaches result in LPV representations with substantial model complexity, both in terms of the number of state variables and the number of scheduling variables. High complexity in LPV models can pose significant challenges for analysis and synthesis, as computational tools can reach hardware limits due to large scheduling dimensions and model orders. Therefore, it is crucial to reduce the complexity of LPV models by minimizing the number of state variables, scheduling variables, but also by reducing the conservativeness of the embedding.

Existing approaches to LPV model reduction have mainly focused on two separate directions: *state-order reduction* (SOR) and *scheduling dimension reduction* (SDR). For SOR, balanced reduction methods, originally developed, for LTI models have been adapted for LPV systems. One example is Moore's *Principal component analysis* (PCA) approach[4]. In the LPV context, this method is extended by incorporating scheduling variations into the static controllability and observability Gramians[5]. Techniques like the balanced reduction for *linear fractional representation* (LFR) models[6] and moment matching for affine LPV-SS models[7] have also been explored. Lastly, a Petrov-Galerkin approximation using parameter-varying oblique projection is introduced in[8]. On the other branch, for SDR, balanced reduction techniques have been adapted to reduce scheduling dimensions by switching the roles of state and latent variables. Later, PCA-based methods became prominent. Kwiatkowski et al.[9] applied PCA to scheduling trajectory data for scheduling dimension reduction, while Sadeghzadeh et al.[10] showed that using data with variations of the state-space matrices directly yields better reduction performance. A kernel-based PCA approach by Rizvi et al.[11] provides an SDR with nonlinear reduction mapping. The capabilities of *neural networks* (NN) have also been investigated in this context, with Rizvi et al.[12] proposing an autoencoder NN for nonlinear reduction mapping. This concept was further extended by Koelewijn et al.[13], which proposes a *deep neural network* (DNN) to directly determine reduced LPV state-space matrices along with scheduling reduction[13].

However, all of the existing methods treat state-order and scheduling dimension reductions as separate problems. While this separation allows for direct solutions, it does not fully exploit the interdependencies between the two reductions. This makes it difficult to simultaneously optimize both aspects of the model, which is crucial for achieving an effective and compact LPV representation. This highlights the potential benefit of a joint reduction methodology, where we address both the state-order and scheduling dimension reductions in a unified framework. Previous work has explored joint reduction approaches, such as the one in[14], where a combination of convex optimization and the LPV Ho-Kalman algorithm was used. However, the computational complexity of this approach limits its practical applicability. Another joint reduction method, developed in[15], specifically targets stable spatially and temporally varying interconnected systems, utilizing the full S-block procedure with both constant and parameter-dependent Gramians to find the appropriate balancing transformation. It requires computing controllability and observability Gramians as multidimensional functions of spatially and temporally varying parameters. This is typically addressed using gridding, which leads to significant computational costs and imprecise results. An additional joint approach is introduced by Beck in[6], where the model is reduced through balanced truncation that simultaneously accounts for state and parameter contributions. This method involves the computation and sorting of the eigenvalues of generalized Hankel operators, which is not a trivial task and can present implementation difficulties, especially in high-dimensional systems.

The multidimensional nature of LPV systems has a natural connection to tensors and tensor-based modeling methods. For instance, tensors helped in addressing the 'curse of dimensionality' in LPV subspace identification[16]. Furthermore, a class of LPV models can be formulated as parameter-varying convex combinations of LTI systems, known as the *Tensor Product* (TP) model[17,18], where tensor decomposition has been successfully used to reduce the number of vertex state-space models representing the overall model variation. Moreover, *proper orthogonal decompositions* (POD) have also been used for lower-rank approximations of multidimensional systems, showcasing the capabilities of tensor methods for data-driven approximations[19].

Driven by these inspirations, the contributions of this paper in solving the joint state and scheduling dimension reduction problem of LPV models are as follows: C1) An extension of the Petrov-Galerkin approximation approach to LPV-SS models with affine dependency, which approach is applicable for both state and scheduling reduction. C2) A novel *tensor-based moment* *matching* (TMM) method for joint reduction of LPV-SS models, where the projection matrices are obtained from reachable and observable subspaces. C3) A data-driven joint *proper orthogonal decomposition* (POD) reduction method of LPV-SS models. C4) Error bounds for the projections of state and scheduling signal using the POD reduction method. C5) A comparative analysis of the proposed joint reduction methods against traditional LPV reduction techniques that independently reduce either the state or the scheduling dimension on challenging examples.

The paper is structured as follows: Section 2 introduces tensors, tensor decompositions, and discrete-time LPV-SS models. Section 3 formulates the problem, focusing on challenges in model complexity and reduction. Section 4 presents the extension of the Petrov-Galerkin approach to LPV-SS models, while Section 5 details the developed tensor-based reduction methods. In Section 6, the benchmark models are detailed on which the advantages of the proposed methods are demonstrated and compared with existing approaches. Section 7 presents the conclusions and open research questions.

## 2 | PRELIMINARIES

This chapter introduces the foundational concepts of tensors and tensor decomposition, followed by definitions of discrete-time LPV-SS representation and key concepts for the reduction of LPV-SS models.

### 2.1 | Tensors

**Definition 1.** An order-$N$ tensor is a multi-linear functional

$$M : \mathcal{W}_1 \times \ldots \times \mathcal{W}_N \to \mathbb{R}$$

acting on inner product spaces[‡] $\mathcal{W}_1, \ldots, \mathcal{W}_N$ with $\langle \cdot, \cdot \rangle_n$ denoting the inner product on $\mathcal{W}_n$, $n = 1, \ldots, N$. In other words, $M$ is a linear functional in each of its $N$ arguments. In the finite dimensional case where $\dim(\mathcal{W}_i) = L_i$, $M$ is referred to as an $L_1 \times L_2 \times \ldots \times L_N$ (covariant) tensor with $L_i$ the $i^{\text{th}}$- mode dimension of $M$. Elements of $M$ are specified by real numbers $m_{l_1,\ldots,l_N}$ where $l_i$ ranges from 1 to $L_i$, and $i$ ranges from 1 to $N$. Elements of $M$ are commonly represented in the $N$-way array $[[m_{l_1,\ldots,l_N}]] \in \mathbb{R}^{L_1 \times \cdots \times L_N}$. The elements $m_{l_1,\ldots,l_N}$ represent $M$ with respect to a specific collection of basis vectors

$$\{e_1^{(l_1)}\}_{l_1=1}^{L_1}, \ldots, \{e_N^{(l_N)}\}_{l_N=1}^{L_N}, \tag{1}$$

for the inner product spaces $\mathcal{W}_1, ..., \mathcal{W}_N$, respectively, in the sense that $m_{l_1,\ldots,l_N} = M(e_1^{(l_1)}, ..., e_N^{(l_N)})$.

The set of all order-$N$ tensors on the inner product spaces $\mathcal{W}_1, \ldots, \mathcal{W}_N$ is denoted by $\mathcal{M}_N$. This set forms a real vector space as it is defined over the field $\mathbb{R}$ when equipped with the standard operations of vector addition and scalar multiplication. Let $\|\cdot\|_n$ represent the norm induced by the inner product $\langle \cdot, \cdot \rangle_n$ on $\mathcal{W}_n$, and it is defined for any vector $w_n$ in $\mathcal{W}_n$ as $\|w_n\|_n = \sqrt{\langle w_n, w_n \rangle_n}$.

First-order tensors are represented as vectors, second-order tensors are represented as matrices, and third and higher-order tensors are represented as $N$-way arrays, which can be written as the following summation:

$$M = \sum_{l_1=1}^{L_1} \cdots \sum_{l_N=1}^{L_N} m_{l_1,\cdots,l_N}\, e_1^{(l_1)} \otimes \cdots \otimes e_N^{(l_N)}, \tag{2}$$

with the *outer product* defined as:

$$\left(e_1^{(l_1)} \otimes \cdots \otimes e_N^{(l_N)}\right)(x_1, \ldots, x_N) = \prod_{n=1}^{N} \langle e_n^{(l_n)}, x_n \rangle_n \tag{3}$$

Each outer product element from the summation (2) we call a *rank-1* tensor. A rank-1 tensor is used as an elementary object to decompose general tensors.

To formulate optimal approximation problems for tensors, a *norm* is introduced on the space $\mathcal{M}_N$.

[‡] As an alternative, the sets $\mathcal{W}_n$ can be taken as general vector spaces together with their algebraic duals $\mathcal{W}_n^*$ with the duality mapping $w_n^* : \mathcal{W}_n \to \mathbb{R}$ denoted as $w_n^*(w_n) = \langle w_n^*, w_n \rangle_n$ for any $w_n^* \in \mathcal{W}_n^*$, $n = 1, \ldots, N$.

**Definition 2.** The operator norm of tensor $M \in \mathcal{M}_N$ is defined as:

$$\|M\|_\circ := \max_{\substack{w_n \in \mathcal{W}_n, \|w_n\|_n = 1 \\ n \in \mathbb{I}_1^N}} \left| M(w_1, \ldots, w_N) \right| \tag{4}$$

with $\mathbb{I}_{\tau_1}^{\tau_2} = \{i \in \mathbb{N} \mid \tau_1 \leq i \leq \tau_2\}$ an index set.

That is, $\|M\|_\circ$ provides the maximal amplitude that a tensor can assume when ranging over the Cartesian product of all unit spheres in $\mathcal{W}_n$, $n \in \mathbb{I}_1^N$. This norm satisfies properties $\|M\|_\circ \geq 0$, $\|M\|_\circ = 0$ only if $M = 0$, $\|\alpha M\|_\circ = |\alpha| \, \|M\|_\circ$ for any scalar $\alpha \in \mathbb{R}$, and $\|M + S\|_\circ \leq \|M\|_\circ + \|S\|_\circ$ for any $M, S \in \mathcal{M}_N$. Therefore, $\mathcal{M}_N$ becomes a normed linear space.

**Definition 3.** The inner product of two tensors $M, S \in \mathcal{M}_N$ with elements $m_{l_1,\ldots,l_N}$ and $s_{k_1,\ldots,k_N}$, both defined with respect to the basis (1), is given by

$$\langle M, S \rangle := \sum_{l_1} \cdots \sum_{l_N} \sum_{k_1} \cdots \sum_{k_N} m_{l_1,\ldots,l_N} s_{k_1,\ldots,k_N} \langle e_1^{(l_1)}, e_1^{(k_1)} \rangle_1 \cdots \langle e_N^{(l_N)}, e_N^{(k_N)} \rangle_N. \tag{5}$$

**Definition 4.** The induced norm of a tensor $M \in \mathcal{M}_N$ is defined as

$$\|M\|_F := \sqrt{\langle M, M \rangle}, \tag{6}$$

and also referred to as Frobenius norm, where $\langle \cdot, \cdot \rangle$ denotes the inner product of tensors in $\mathcal{M}_N$ defined in Definition 3.

Tensor contraction refers to the operation of summing over one or more common indices between tensors. This operation generalizes matrix multiplication and plays an essential role in many tensor decompositions and algorithms. Two important forms of tensor contraction used in this work are introduced below.

**Definition 5.** (Tensor-matrix multiplication: The $n$-mode product) The $n$-mode product of a tensor $M \in \mathbb{R}^{L_1 \times L_2 \times \cdots \times L_N}$, having elements $m_{l_1,\ldots,l_N}$, with a matrix $Q \in \mathbb{R}^{R_n \times L_n}$, having elements $q_{r_n,l_n}$, is denoted by $M \cdot_n Q$ and is of size $L_1 \times \cdots \times L_{n-1} \times R_n \times L_{n+1} \times \cdots \times L_N$. Element-wise, it can be written as:

$$(M \cdot_n Q)_{l_1,l_2,\ldots,l_{n-1},r_n,l_{n+1},\ldots,l_N} = \sum_{l_n=1}^{L_n} m_{l_1 l_2,\ldots,l_N} \, q_{r_n,l_n},$$

**Definition 6.** (Tensor-Tensor Multiplication) Let $M \in \mathbb{R}^{L_1 \times L_2 \times \cdots \times L_M}$ and $S \in \mathbb{R}^{J_1 \times J_2 \times \cdots \times J_N}$ be two tensors. We define the *tensor-tensor multiplication* along one mode from each tensor as the summation over the common dimension. Let $L_i$ and $J_k$ be the modes to be contracted, where $L_i = J_k$. The tensor-tensor product $(*_{i,j})$ between $M$ and $S$ produces a tensor:

$$(M *_{i,k} S) : (L_1 \times \cdots \times L_{i-1} \times L_{i+1} \times \cdots \times L_M \times J_1 \times \cdots \times J_{k-1} \times J_{k+1} \times \cdots \times J_N) \to \mathbb{R}$$

with elements:

$$(M *_{i,k} S)_{l_1,\ldots,l_{i-1},l_{i+1},\ldots,l_M,j_1,\ldots,j_{k-1},j_{k+1},\ldots,j_N} = \sum_{l=1}^{L_i} m_{l_1,\ldots,l_{i-1},l,l_{i+1},\ldots,l_M} s_{j_1,\ldots,j_{k-1},l,j_{k+1},\ldots,j_N}$$

The concept of *tensor rank* is a non-trivial extension of the same concept for linear mappings and has been discussed in detail in, e.g.[20]. The rank of a tensor provides insight into its complexity and the minimal number of rank-1 tensors required to represent it. This concept is analogous to the matrix rank, but extends to higher-dimensional tensors.

**Definition 7.** (Tensor rank) The rank of a tensor $M \in \mathcal{M}_N$, denoted as rank($M$), is defined as the minimum integer $R$ such that $M$ can be decomposed into a sum of $R$ rank-1 tensors.

To further understand tensor rank, we introduce the notion of *n-mode rank*. This concept involves analyzing the tensor with respect to its modes (or dimensions) individually.

**Definition 8.** (The $n$-mode kernel) For a tensor $M \in \mathcal{M}_N$, the *n-mode kernel* is defined as

$$\ker_n(M) := \left\{ w_n \in \mathcal{W}_n \;\middle|\; M(w_1, \ldots, w_N) = 0, \; \forall w_i \in \mathcal{W}_i, \; i \neq n \right\}. \tag{7}$$

This set consists of all vectors in the $n$-th mode that nullify the tensor when combined with any choice of vectors in the other modes.

**Definition 9.** The *n-mode rank* of a tensor $M \in \mathcal{M}_N$ is defined by

$$R_n = \text{rank}_n(M) := \dim(\mathcal{W}_n) - \dim(\ker_n(M)), \quad n \in \mathbb{I}_1^N. \tag{8}$$

This measure provides insight into the dimensionality of the $n$-th mode, considering the null space of the tensor in that mode. Additionally, $n$–mode rank coincides with the rank of a matrix obtained by unfolding tensor $M$ in dimension $n$.

**Definition 10.** (Modal rank) The *modal rank* or *multi-linear rank* of a tensor $M \in \mathcal{M}_N$ , denoted as modrank$(M)$, is a vector consisting of all the $n$-mode ranks:

$$\text{modrank}(M) = (R_1, R_2, \ldots, R_N), \tag{9}$$

with $R_n = \text{rank}_n(M)$.

## 2.2 | Tensor decomposition

A general overview of tensor decomposition is presented, followed by a detailed discussion of *tensor singular value decomposition* (TSVD) and *higher-order singular value decomposition* (HOSVD).

### 2.2.1 | General concepts

**Definition 11.** Consider a tensor $M \in \mathcal{M}_N$ that operates on a collection of vector spaces, $M : \mathcal{W}_1 \times \cdots \times \mathcal{W}_N \to \mathbb{R}$. By changing the basis of $\mathcal{W}_n$ of the representation of $M$ with respect to new basis vectors $\{\phi_n^{(l_n)}\}_{l_n=1}^{L_n}$, where $n \in \mathbb{I}_1^N$. In the new basis, the tensor $M$ is represented as:

$$M = \sum_{l_1=1}^{L_1} \cdots \sum_{l_N=1}^{L_N} m_{l_1 \cdots l_N} \, \phi_1^{(l_1)} \otimes \cdots \otimes \phi_N^{(l_N)}, \tag{10}$$

where the $N$-way array $[[m_{l_1,\cdots,l_N}]]$ is called the *core tensor*.

Every single term inside of the sum (10) with an outer product of $N$ vectors represents a rank-1 tensor. The tensor $M$ is formed as a weighted sum of rank-1 tensors. Tensor decomposition essentially depends on a change of basis vectors that are used to represent the tensor $M$. The desirable properties that we aim to achieve with tensor decompositions include:

1. **Diagonality**: The core tensor in the representation (10) is diagonal, meaning $m_{l_1 \cdots l_N} = 0$ unless $l_1 = \cdots = l_N$.
2. **Orthonormality**: The decomposition (10) often satisfies one of the following orthonormality properties:
   (a) **Completely orthonormal decomposition**: For each $n \in \mathbb{I}_1^N$, the set $\{\phi_n^{(l_n)}\}_{l_n=1}^{L_n}$ forms an orthonormal basis of $\mathcal{W}_n$ with respect to $\langle \cdot, \cdot \rangle_n$.
   (b) **Orthonormal (orthogonal) decomposition**: The terms $(\phi_1^{(l_1)} \otimes \cdots \otimes \phi_N^{(l_N)})$ from (10) are mutually orthonormal (orthogonal) for all index combinations $(l_1, \ldots, l_N)$, with respect to the inner product $\langle \cdot, \cdot \rangle$ of the tensor space $\mathcal{M}_N$.
3. **Low Approximation Error**: The decomposition (10) of $M$ allows for approximations of $M$. Truncations in terms of the number of basis in the summations generally lead to lower rank approximations $\hat{M}$ where $\|M - \hat{M}\|_F$ is relevant.

For matrices, the singular value decomposition satisfies all three properties, where the lower-rank approximation provides the optimal result in terms of minimizing the $(2, 2)$–norm of the error. However, for tensors, it is not always possible to construct a tensor decomposition that satisfies all three properties[21]. A detailed overview of tensor decompositions with different computational methods is provided in[22].

### 2.2.2 | Singular value decomposition for tensors

One crucial decomposition for model reduction is the *tensor singular value decomposition* (TSVD)[21]. Let $M \in \mathcal{M}_N$ be an order-$N$ tensor defined on the finite dimensional vector spaces $\mathcal{W}_1, \ldots, \mathcal{W}_N$, where $\dim(\mathcal{W}_n) = L_n$, for all $n \in \mathbb{I}_1^N$. The *singular values* of $M$, denoted as $\sigma_i(M)$ with $i \in \mathbb{I}_1^R$, where $R = \min(\text{modrank}(M))$, are defined as follows.

For $n \in \mathbb{I}_1^N$, let

$$\mathcal{S}_n^{(1)} := \{\phi \in \mathcal{W}_n \mid \|\phi\|_n = 1\} \tag{11}$$

denote the unit sphere in $\mathcal{W}_n$. The first singular value of $M$ is defined as:

$$\sigma_1(M) := \sup_{\phi_n \in \mathcal{S}_n^{(1)},\ 1 \le n \le N} \left| M(\phi_1, \ldots, \phi_N) \right|. \tag{12}$$

As the Cartesian product $\mathcal{S}^{(1)} = \mathcal{S}_1^{(1)} \times \ldots \times \mathcal{S}_N^{(1)}$ of unit spheres is a compact set and $M$ is continuous, an extremal solution of (12) exists and is attained by the $N$-tuple

$$\left(\phi_1^{(1)}, \ldots, \phi_N^{(1)}\right) \in \mathcal{S}^{(1)}.$$

The subsequent singular values of $M$ are defined inductively via the orthogonality condition:

$$\mathcal{S}_n^{(i)} := \left\{ \phi \in \mathcal{W}_n \mid \|\phi\|_n = 1,\ \langle \phi, \phi_n^{(j)} \rangle = 0 \ \text{ with } \phi_n^{(j)} \in \mathcal{S}_n^{(j)} \text{ for } j \in \mathbb{I}_1^{i-1} \right\},$$

for $n \in \mathbb{I}_1^N$, $i \geq 1$, providing

$$\sigma_i(M) := \sup_{\phi_n \in \mathcal{S}_n^{(i)},\ 1 \le n \le N} \left| M(\phi_1, \ldots, \phi_N) \right|. \tag{13}$$

The obtained vectors $\phi_n^{(1)}, \ldots, \phi_n^{(R)}$, $n = 1, \ldots, N$, are mutually orthonormal in the vector space $\mathcal{W}_n$. If $R < L_n$ for any $n$, then the collection of orthonormal elements is extended to a complete orthonormal basis of $\mathcal{W}_n$. This leads to a collection of orthonormal bases $\phi_n(M) = \{\phi_n^{(l_n)}\}_{l_n=1}^{L_n}$ for $\mathcal{W}_n$, where $n \in \mathbb{I}_1^N$. A method to compute the tensor singular values and vectors is proposed in[21], and the algorithm is based on the fixed point properties of contraction mapping. For the singular values $\sigma_1(M), \ldots, \sigma_R(M)$, the natural ordering in terms of magnitude is considered, i.e.,

$$\sigma_1(M) \geq \sigma_2(M) \geq \cdots \geq \sigma_R(M) > 0.$$

The next theorem from[21] states that for *diagonalizable* tensors, the TSVD provides an optimal lower-rank approximation.

**Definition 12.** A tensor $M \in \mathcal{M}_N$ is diagonalizable whenever a collection of bases can be found such that (10) has diagonal core, i.e. $m_{l_1,\ldots,l_n} \neq 0$ only if $l_1 = \cdots = l_N$.

**Theorem 1.** *Let the tensor $M \in \mathcal{M}_N$ be diagonalizable with respect to an orthonormal basis. Then, the singular value decomposition of $M$ is a completely orthogonal rank decomposition. Moreover, the truncated tensor $M_r^*$ represented as:*

$$M_r^* = \sum_{i=1}^{r} \sigma_i \phi_1^{(i)} \otimes \cdots \otimes \phi_N^{(i)}, \tag{14}$$

*is an optimal rank-r approximation of $M$ in the sense that:*

$$\inf_{M_r \in \mathcal{M}_N,\ \text{rank}(M_r) = r} \|M - M_r\|_\circ = \|M - M_r^*\|_\circ = \sigma_{r+1}, \tag{15}$$

*where $\|\cdot\|$ is the operator norm of a tensor. Moreover,*

$$\inf_{M_r \in \mathcal{M}_N,\ \text{rank}(M_r) = r} \|M - M_r\|_F^2 = \|M - M_r^*\|_F^2 = \sum_{n=r+1}^{R} \sigma_n^2, \tag{16}$$

*where $\|\cdot\|_F$ is the Frobenius norm of the tensor.*

Proof of Theorem 1 can be found in[21].

### 2.2.3 | Higher-order singular value decomposition

A special widely used tensor decomposition is the *higher-order singular value decomposition* (HOSVD)[23]. Given a tensor $M \in \mathcal{M}_N$ with $\mathrm{modrank}(M) = (R_1, \ldots, R_N)$. Then, tensor $M$ can be represented as the multilinear tensor-matrix product of a core tensor $S$ of size $R_1 \times R_2 \times \ldots \times R_N$ with $N$ matrices $Q_n$ of size $L_n \times R_n$:

$$M = S \cdot_1 Q_1 \cdot_2 Q_2 \cdot_3 \ldots \cdot_N Q_N, \tag{17}$$

where $\cdot_n$ denotes a tensor-matrix multiplication in mode $n$. HOSVD can be computed by finding singular vectors of the unfolded tensor along the dimension $n$. HOSVD falls under the larger category of tensor decompositions called Tucker decompositions, for more details, see[23].

The Tucker decomposition allows finding a lower-rank approximation of any chosen modal rank of a tensor $\mathrm{modrank}(M) = (r_1, \ldots, r_N)$, or *reduced* HOSVD, by preserving only $r_i$ columns from the matrix $Q_i$, for $i \in \mathbb{I}_1^N$. The lower-rank approximation of tensor $M$ can be represented in a form:

$$M_r = \sum_{l_1=1}^{r_1} \ldots \sum_{l_N=1}^{r_N} s_{l_1,\ldots,l_N}\, \phi_1^{(l_1)} \otimes \ldots \otimes \phi_N^{(l_N)}, \tag{18}$$

where $s_{l_1,\ldots,l_N}$ are the elements from the core tensor $S$, and $\phi_i^{l_i}$ is $l_i$−th column vector from the matrix $Q_i$, which is also a singular vector.

## 2.3 | Discrete-time LPV-SS models

Consider a *discrete-time* (DT) LPV-SS representation $\Sigma$:

$$\Sigma := \begin{cases} x(t+1) = A(p(t))x(t) + B(p(t))u(t) \\ \quad\;\; y(t) = C(p(t))x(t) + D(p(t))u(t) \end{cases} \tag{19}$$

where $t \in \mathbb{Z}_0^+$ is a discrete-time instance, $x(t) \in \mathbb{X} \subseteq \mathbb{R}^{n_\mathrm{x}}$ is a state variable, $u(t) \in \mathbb{U} \subseteq \mathbb{R}^{n_\mathrm{u}}$ is the input, and $y(t) \in \mathbb{Y} \subseteq \mathbb{R}^{n_\mathrm{y}}$ is the output, and $p(t) \in \mathbb{P} \subset \mathbb{R}^{n_\mathrm{p}}$ is the scheduling variable, and there exists a function $\eta(x(t), u(t)) = p(t)$ called the scheduling map which describes the internal dependence of scheduling variables on state and input values. An LPV-SS representation $\Sigma$ is driven by the input $u = \{u(t)\}_{t=0}^{\infty}$ and the scheduling sequence $p = \{p(t)\}_{t=0}^{\infty}$. Different types of dependence of system matrices on the scheduling variable can be defined, such as affine, polynomial, and rational. Further, let us assume that the matrix functions $A, B, C, D$ are affinely dependent on the scheduling variables $p(t)$. Merging the affine term in a way $\bar{p}(t) = \begin{bmatrix} p_0(t) & p_1(t) & \ldots & p_{n_\mathrm{p}}(t) \end{bmatrix}^T$, with $p_0(t) = 1$ for all $t \in \mathbb{Z}_0^+$, allows us to rewrite the matrix functions using order-3 tensors:

$$A\big(p(t)\big) = \sum_{i=0}^{n_\mathrm{p}} p_i(t) A_i = \mathcal{A} \cdot_3 \bar{p}^\top(t), \quad B\big(p(t)\big) = \sum_{i=0}^{n_\mathrm{p}} p_i(t) B_i = \mathcal{B} \cdot_3 \bar{p}^\top(t), \tag{20a}$$

$$C\big(p(t)\big) = \sum_{i=0}^{n_\mathrm{p}} p_i(t) C_i = \mathcal{C} \cdot_3 \bar{p}^\top(t), \quad D\big(p(t)\big) = \sum_{i=0}^{n_\mathrm{p}} p_i(t) D_i = \mathcal{D} \cdot_3 \bar{p}^\top(t). \tag{20b}$$

where $\mathcal{A} \in \mathbb{R}^{n_\mathrm{x} \times n_\mathrm{x} \times (n_\mathrm{p}+1)}$, $\mathcal{B} \in \mathbb{R}^{n_\mathrm{x} \times n_\mathrm{u} \times (n_\mathrm{p}+1)}$, $\mathcal{C} \in \mathbb{R}^{n_\mathrm{y} \times n_\mathrm{x} \times (n_\mathrm{p}+1)}$, $\mathcal{D} \in \mathbb{R}^{n_\mathrm{y} \times n_\mathrm{u} \times (n_\mathrm{p}+1)}$. Then, (19) can be represented in tensor form as

$$\Phi := \begin{cases} x(t+1) = \mathcal{A} \cdot_2 x^\top(t) \cdot_3 \bar{p}^\top(t) + \mathcal{B} \cdot_2 u^\top(t) \cdot_3 \bar{p}^\top(t) \\ \quad\;\; y(t) = \mathcal{C} \cdot_2 x^\top(t) \cdot_3 \bar{p}^\top(t) + \mathcal{D} \cdot_2 u^\top(t) \cdot_3 \bar{p}^\top(t) \end{cases}. \tag{21}$$

Now, we introduce the concepts from the realization theory for LPV-SS representations of the form of (19) that are essential for the sequel. For more details see[24]. We denote by $H^{\mathbb{N}}$ the set of all maps of the form $f : \mathbb{N} \to H$ where $H$ is a (possibly infinite) set. Then, the sets $\mathcal{U}$, $\mathcal{P}$, $\mathcal{Y}$, $\mathcal{X}$ are defined as $\mathcal{U} = \mathbb{U}^{\mathbb{N}}$, $\mathcal{P} = \mathbb{P}^{\mathbb{N}}$, $\mathcal{Y} = \mathbb{Y}^{\mathbb{N}}$, and $\mathcal{X} = \mathbb{X}^{\mathbb{N}}$.

Consider an initial state $x_0 \in \mathbb{R}^{n_\mathrm{x}}$ of the LPV-SS representation $\Sigma$ of the form (19). The *input-to-state map* $X_{\Sigma,x_0} : \mathcal{U} \times \mathcal{P} \to \mathcal{X}$ and *input-to-output (I/O) map* $Y_{\Sigma,x_0} : \mathcal{U} \times \mathcal{P} \to \mathcal{Y}$ of $\Sigma$ corresponding to this initial state $x_0$ are defined as follows: for all sequences $u = \{u(t)\}_{t=0}^{\infty} \in \mathcal{U}$ and $p = \{p(t)\}_{t=0}^{\infty} \in \mathcal{P}$, let $X_{\Sigma,x_0}(u,p)(t) = x(t)$ and $Y_{\Sigma,x_0}(u,p)(t) = y(t)$, where $x(t)$, $y(t)$ satisfy (19) with $x(0) = x_0$, and $t \in \mathbb{Z}_0^+$.

**Definition 13.** (Reachability of an LPV-SS representation) We say that $\Sigma$ is *reachable*, if $\mathbb{R}^{n_x} = \text{span}\{X_{\Sigma,0}(u,p)(t) \mid (u,p) \in \mathcal{U} \times \mathcal{P},\ t \in \mathbb{Z}_0^+\}$, i.e., $\mathbb{R}^{n_x}$ is the smallest vector space containing all the states which are reachable from $x(0) = 0$ by some scheduling sequence and input sequence at some time instance $t$, where $t \in \mathbb{Z}_0^+$.

**Definition 14.** (Observability of an LPV-SS representation) We say that $\Sigma$ is *observable* if for any two initial states $x_1, x_2 \in \mathbb{R}^{n_x}$, $Y_{\Sigma,x_1} = Y_{\Sigma,x_2}$ implies $x_1 = x_2$. That is, if any two distinct initial states of an observable $\Sigma$ are chosen, then for *some* input and scheduling sequence, the resulting outputs will be different.

**Definition 15.** (I/O equivalence of LPV-SS representations) Two LPV-SS representations $\Sigma_1$ and $\Sigma_2$ are called *input-output (I/O) equivalent* if $Y_{\Sigma_1,0} = Y_{\Sigma_2,0}$.

Note that two LPV-SS representations can be I/O equivalent even in the case where they have a different number of state variables ($\dim_x(\Sigma)$) and a different number of scheduling variables ($\dim_p(\Sigma)$).

In fact, the input-output behavior of an LPV-SS representation $\Sigma$ (19) can be formalized as a map $f : \mathcal{U} \times \mathcal{P} \to \mathcal{Y}$, independent from the actual choice of the state basis. The value $f(u,p)(t)$ represents the output of the system at time $t$, starting from zero initial conditions, when the input sequence $u = \{u(t)\}_{t=0}^{\infty}$, and the scheduling sequence $p = \{p(t)\}_{t=0}^{\infty}$ are fed to the system. The LPV-SS representation of the form (19) is a realization of the map $f$, if $f$ equals the input-output map of $\Sigma$ for $x_0 = 0$, i.e. $f = Y_{\Sigma,0}$.

**Definition 16.** (Minimal realization of I/O map) An LPV-SS realization $\Sigma$ is a minimal realization of the I/O map $f$ if $\Sigma$ is a realization of $f$, and for any LPV-SS representation $\bar{\Sigma}$ which is also a realization of $f$, $\dim_x(\Sigma) \leq \dim_x(\bar{\Sigma})$ and $\dim_p(\Sigma) \leq \dim_p(\bar{\Sigma})$. We say that $\Sigma$ is *minimal* if $\Sigma$ is a minimal realization of its own input-output map $Y_{\Sigma}$.

The I/O map $f$ can be represented as an *infinite impulse response* (IIR) representation. Consider an LPV-SS representation $\Sigma$ of the form (19), and consider its I/O map $f = Y_{\Sigma,0}$. For any input sequence $u = \{u(t)\}_{t=0}^{\infty}$ and scheduling sequence $p = \{p(t)\}_{t=0}^{\infty}$, the IIR form of $f = Y_{\Sigma,0}$ is given as

$$f(u,p)(t) = Y_{\Sigma,0}(u,p)(t) = \sum_{\tau=0}^{t} (h_\tau \diamond p)(t) u(t-\tau), \tag{22}$$

for all $t \in \mathbb{N}$, where

$$(h_0 \diamond p)(t) = D(p(t)), \quad (h_1 \diamond p)(t) = C(p(t))B(p(t-1)), \quad \ldots \quad , (h_m \diamond p)(t) = C(p(t))\Big(\prod_{l=1}^{m-1} A(p(t-l))\Big)B(p(t-m)).$$

## 3 | PROBLEM FORMULATION

The main objective of this paper is to approximate a discrete-time LPV-SS model $\Sigma$, described by (19), with a reduced-order model $\Sigma_r$ that minimizes the number of states and scheduling variables while preserving the essential I/O dynamics of the full-order model. We will develop a solution for this problem using tensors, which will allow to reduce both the states and scheduling variables simultaneously, while providing flexibility on the complexity/accuracy trade-off.

The reduced-order affine LPV-SS model $\Sigma_r$ is considered as

$$\Sigma_r := \begin{cases} x_r(t+1) = A_r(p_r(t))x_r(t) + B_r(p_r(t))u(t) \\ \quad\ y_r(t) = C_r(p_r(t))x_r(t) + D_r(p_r(t))u(t) \end{cases} \tag{23}$$

with reduced state-dimension $\dim(x_r) = r_x \leq n_x$ and scheduling dimension $\dim(p_r) = r_p \leq n_p$. In (23), $x_r(t) \in \mathbb{R}^{r_x}$ is the reduced state variable, $p_r(t) \in \mathbb{R}^{r_p}$ is reduced scheduling variable, and the matrix functions $A_r, B_r, C_r$, and $D_r$ are affinely dependent on $p_r$, together with a reduced-dimension scheduling map:

$$p_r(t) = \eta_r(x_r(t), u(t)). \tag{24}$$

The reduced LPV-SS model $\Sigma_r$ can be represented in a tensor form as:

$$\Phi_r := \begin{cases} x_r(t+1) = \mathcal{A}_r \cdot_2 x_r^\top(t) \cdot_3 p_r^\top(t) + \mathcal{B}_r \cdot_2 u^\top(t) \cdot_3 p_r^\top(t) \\ \quad\; y(t) = \mathcal{C}_r \cdot_2 x_r^\top(t) \cdot_3 p_r^\top(t) + \mathcal{D}_r \cdot_2 u^\top(t) \cdot_3 p_r^\top(t) \end{cases} \tag{25}$$

where tensors $\mathcal{A}_r, \mathcal{B}_r, \mathcal{C}_r, \mathcal{D}_r$ completely define the reduced-order model $\Phi_\mathrm{r}$.

Next we will formulate our approach for joint reduction of state and scheduling variables of a discrete-time LPV-SS representation, discussing how $r_\mathrm{x}$ and $r_\mathrm{p}$ are determined, in the following sections.

## 4 | EXTENSION OF THE PETROV-GALERKIN PROJECTION FOR LPV-SS MODELS

A common projection approach used for model approximation is the Petrov-Galerkin projection[25]. This approach consists of two fundamental projections: a state projection and the residual (test space) projection.

The original system state evolves in the function space $\mathcal{X} := \left\{x : \mathbb{Z}_0^+ \to \mathbb{R}^{n_\mathrm{x}}\right\}$, which consists of time-dependent signals taking values in the state-space $\mathbb{X} = \mathbb{R}^{n_\mathrm{x}}$. The state projection is defined by $x \approx Vx_\mathrm{r}$ where $V \in \mathbb{R}^{n_\mathrm{x} \times r_\mathrm{x}}$ is the state projection matrix. The approximation of the state $\hat{x}$ resides in $\mathrm{im}(V) = \mathcal{V}$, assuming w.l.o.g. that $V$ is full rank. The reduced state space is then $\mathcal{X}_\mathrm{r} := \left\{x_\mathrm{r} : \mathbb{Z}_0^+ \to \mathbb{R}^{r_\mathrm{x}}\right\}$. Every state vector $x(t) \in \mathbb{X}$ can be decomposed as:

$$x(t) = \underbrace{\hat{x}(t)}_{\in \mathcal{V}} + \underbrace{\tilde{x}(t)}_{\in \mathcal{V}^\perp}, \tag{26a}$$

$$\hat{x}(t) = \Pi_\mathcal{V} x(t) = V\underbrace{(V^\top V)^{-1} V^\top x(t)}_{x_\mathrm{r}(t)} = Vx_\mathrm{r}(t), \tag{26b}$$

$$x_\mathrm{r}(t) = (V^\top V)^{-1} V^\top x(t) = V^\dagger x(t), \tag{26c}$$

where $\Pi_\mathcal{V}$ denotes the orthogonal projection onto $\mathcal{V}$, and $V^\dagger$ denotes the Moore-Penrose inverse of $V$.

Since the projected state $x_\mathrm{r}(t)$ must still satisfy a reduced-order version of the system equations, we can enforce a second projection on the residual. The residual is defined as the error introduced by replacing $x(t)$ with its projected counterpart in the system equations. To ensure a well-posed reduced system, we require that this residual is orthogonal to a chosen test space, spanned by a matrix $W \in \mathbb{R}^{n_\mathrm{x} \times r_\mathrm{x}}$. Thus, the test space projection (or residual projection) guarantees that the reduction does not introduce uncontrolled approximation errors. A common selection is $W = V$, which results in a Galerkin projection (a special case where the test space is identical to the trial space). However, in Petrov-Galerkin projection, $W$ is chosen independently to optimize the reduction.

As one of the main contributions of this paper, the Petrov-Galerkin projection is extended for LPV-SS representations by including a third projection on the scheduling signal. Similarly to the state projection, the scheduling projection is defined by $p \approx Zp_\mathrm{r}$ where $Z \in \mathbb{R}^{n_\mathrm{p} \times r_\mathrm{p}}$ is the scheduling projection matrix. The approximation of the scheduling variables $\hat{p}$ resides in $\mathrm{im}(Z) = \mathcal{Z}$, assuming w.l.o.g that $Z$ is full rank. The reduced scheduling space is then $\mathcal{P}_\mathrm{r} := \left\{p_r : \mathbb{Z}_0^+ \to \mathbb{R}^{r_\mathrm{p}}\right\}$. Similarly as before, every scheduling vector $p(t) \in \mathbb{P}$ can be decomposed as:

$$p(t) = \underbrace{\hat{p}(t)}_{\in \mathcal{Z}} + \underbrace{\tilde{p}(t)}_{\in \mathcal{Z}^\perp}, \tag{27a}$$

$$\hat{p}(t) = \Pi_\mathcal{Z}\, p(t) = Z\underbrace{(Z^\top Z)^{-1} Z^\top p(t)}_{p_\mathrm{r}(t)} = Zp_\mathrm{r}(t), \tag{27b}$$

$$p_\mathrm{r}(t) = (Z^\top Z)^{-1} Z^\top p(t) = Z^\dagger p(t), \tag{27c}$$

where $\Pi_\mathcal{Z}$ refers to the orthogonal projection onto $\mathcal{Z}$ and $Z^\dagger$ refers to the Moore-Penrose inverse of $Z$.

State, residual, and scheduling projections, defined by the matrices $V$, $W$, and $Z$ completely describe the LPV-SS projection to the lower-order model and they represents the core mechanism of LPV-SS model reduction. Now, we will apply the three projections on the discrete-time LPV-SS model $\Sigma$ of the form (19) to obtain general forms of reduced models.

**Algorithm 1** Petrov-Galerkin projection for affine LPV-SS representations

**Inputs:** discrete-time LPV model with affine dependency $\Sigma$, projection matrices $V$, $W$, $Z$

**Output:** $\Sigma_r = (\mathcal{A}_\mathrm{r}, \mathcal{B}_\mathrm{r}, \mathcal{C}_\mathrm{r}, \mathcal{D}_\mathrm{r}, \eta_\mathrm{r})$

1: **construct:** tensor form of the LPV-SS representation $\Sigma$ from (21) as $(\mathcal{A}, \mathcal{B}, \mathcal{C}, \mathcal{D}, \eta)$

2: **calculate:** $(\mathcal{A}_\mathrm{r}, \mathcal{B}_\mathrm{r}, \mathcal{C}_\mathrm{r}, \mathcal{D}_\mathrm{r})$ based on (28)

3: **calculate:** the reduced scheduling map $\eta_\mathrm{r}$ based on (29) or (30)

4: **return:** $\Sigma_r = (\mathcal{A}_\mathrm{r}, \mathcal{B}_\mathrm{r}, \mathcal{C}_\mathrm{r}, \mathcal{D}_\mathrm{r}, \eta_\mathrm{r})$

Looking from the perspective of the tensorial form of the LPV-SS representation, given by (21), the first step is to enforce the orthogonality of the residual to the chosen test-space defined by $W$:

$$\left\langle x(t+1) - \mathcal{A} \cdot_2 x^\top(t) \cdot_3 p^\top(t) - \mathcal{B} \cdot_2 u^\top(t) \cdot_3 p^\top(t),\ \xi \right\rangle = 0,$$

for every $\xi \in \mathcal{W}$, $x(t) \in \mathbb{X}$, $p(t) \in \mathbb{P}$, and $u(t) \in \mathbb{U}$. Next, the approximations of the state signal $\hat{x}(t)$ of $x(t)$ and the scheduling signal $\hat{p}(t)$ of $p(t)$ are included in the equation above with

$$\begin{aligned}
&\left\langle \hat{x}(t+1) - \mathcal{A} \cdot_2 \hat{x}^\top(t) \cdot_3 \hat{p}^\top(t) - \mathcal{B} \cdot_2 u^\top(t) \cdot_3 \hat{p}^\top(t),\ Ww \right\rangle = 0, \quad \forall w \in \mathbb{R}^{r_\mathrm{x}},\\
&\left\langle W^\top V x_r(t+1) - W^\top \mathcal{A} \cdot_2 (V x_\mathrm{r}(t))^\top \cdot_3 (Z p_\mathrm{r}(t))^\top - W^\top \mathcal{B} \cdot_2 u^\top(t) \cdot_3 (Z p_\mathrm{r}(t))^\top,\ w \right\rangle = 0, \quad \forall w \in \mathbb{R}^{r_\mathrm{x}},\\
&W^\top V x_r(t+1) = W^\top \mathcal{A} \cdot_2 (V x_\mathrm{r}(t))^\top \cdot_3 (Z p_\mathrm{r}(t))^\top + W^\top \mathcal{B} \cdot_2 u^\top(t) \cdot_3 (Z p_\mathrm{r}(t))^\top.
\end{aligned}$$

Finally, the reduced LPV-SS model $\Phi_r$ from (25) is described by the following state and output equations:

$$x_r(t+1) = \underbrace{\left(\mathcal{A} \cdot_1 (W^\top V)^{-1} W^\top \cdot_2 V^\top \cdot_3 Z^\top\right)}_{\mathcal{A}_\mathrm{r}} \cdot_2 x_\mathrm{r}^\top(t) \cdot_3 p_\mathrm{r}^\top(t) + \underbrace{\left(\mathcal{B} \cdot_1 (W^\top V)^{-1} W^\top \cdot_3 Z^\top\right)}_{\mathcal{B}_\mathrm{r}} \cdot_2 u^\top(t) \cdot_3 p_\mathrm{r}^\top(t) \tag{28a}$$

$$y_r(t) = \underbrace{\left(\mathcal{C} \cdot_2 V^\top \cdot_3 Z^\top\right)}_{\mathcal{C}_\mathrm{r}} \cdot_2 x_\mathrm{r}^\top(t) \cdot_3 p_\mathrm{r}^\top(t) + \underbrace{\left(\mathcal{D} \cdot_3 Z^\top\right)}_{\mathcal{D}_\mathrm{r}} \cdot_2 u^\top(t) \cdot_3 p_\mathrm{r}^\top(t), \tag{28b}$$

together with the scheduling map of the approximated model, from (24):

$$\eta_\mathrm{r}(x_\mathrm{r}(t), u(t)) = p_\mathrm{r}(t) = (Z^\top Z)^{-1} Z^\top \hat{p}(t) = Z^\dagger \eta(V x_\mathrm{r}(t), u(t)). \tag{29}$$

We will see that the construction of the scheduling space $\mathcal{Z}$ varies in the developed methods in a sense that it can also include the affine term associated $p_0(t) = 1$ inside of the scheduling vector, enforcing the transformation $Z : \mathbb{R}^{n_\mathrm{p}+1} \to \mathbb{R}^{r_\mathrm{p}}$. The affine term from the LPV-SS model is then transformed, and the scheduling map becomes:

$$p_\mathrm{r}(t) = \eta_\mathrm{r}(x_\mathrm{r}(t), u(t)) = Z^\dagger \begin{pmatrix} 1 \\ \eta(V x_\mathrm{r}(t), u(t)) \end{pmatrix}. \tag{30}$$

The above derivation shows how to project an affine LPV-SS model to a lower-order subspace and reduce both the number of states and scheduling variables. The LPV-SS Petrov-Galerkin projection is summarized in Algorithm 1. The remaining question is how to choose projection matrices $V$, $W$, and $Z$, or rather, how to find efficiently subspaces that capture well the variation of the state and scheduling trajectories.

Note that the introduced projection allows an independent projection of the state, scheduling and residual signals. The common choice for residual projection is to enforce $W = V$, which is motivated by the idea to have both state variables and the derivatives of state variables in the same subspace. This type of projection that enforces orthogonality is called a Galerkin projection.

# 5 | TENSOR-BASED LPV MODEL REDUCTION METHODS

As our second and third contributions, we present novel methods for joint state-order and scheduling signal dimension reduction of LPV systems. These methods leverage tensor decomposition to obtain a reduced state space $\mathbb{X}_\mathrm{r}$ and a reduced scheduling space $\mathbb{P}_\mathrm{r}$. With these methods, we obtain the projection matrices $V$, $W$, and $Z$, which define the lower-dimensional subspaces for states, residual, and scheduling signals. Then, we use them in the LPV Petrov-Galerkin setting introduced in Section 4 to obtain the reduced model.

The core reduction methods we develop are the *tensor-based LPV moment matching* (TMM) and the *proper orthogonal decomposition* (POD) approach. These methods differ in how they construct the reduced subspaces $\mathbb{X}_\mathrm{r}$ and $\mathbb{P}_\mathrm{r}$. The tensor-based LPV moment matching method builds the reduced subspaces based on the matrix function of an existing LPV-SS representation. In contrast, the tensor-based POD method constructs the reduced subspaces using simulation data containing state and scheduling trajectories, making it applicable even when an analytic form of the LPV model is unavailable. This distinction highlights the flexibility of the tensor-based POD method in data-driven scenarios, while TMM remains closely tied to the structure of the original LPV model. As an important result, we will also show that *reachability* and *observability* subspaces of affine LPV models can be defined not only for state variables, but also for the scheduling variables, which are crucial for the tensor-based moment matching method.

## 5.1 | Tensor-based LPV moment matching

The extension of moment matching for discrete LPV-SS representations affinely dependent on the scheduling variables has been introduced in[7]. It preserves the $n$–partial input-output map, making the $n$–partial I/O map of the full-order model $\Sigma$ equivalent to the $n$–partial I/O map of the reduced-order model $\Sigma_r$. Out objective is to find an extension of moment matching that allows us to reduce the state and scheduling variables simultaneously. We used tensors to achieve this goal.

Firstly, let us introduce the *reachability tensors* of an LPV-SS tensor-representation $\Sigma$ described in (21). Let $\mathcal{A}$ and $\mathcal{B}$ be the state transition and input tensors. Then, the reachability tensors for horizon $n$ can be obtained recursively as:

$$\begin{aligned}
\mathcal{R}_0 &= \tilde{\mathcal{B}}, && \dim(\mathcal{R}_0) = (n_\mathrm{x}, n_\mathrm{p}+1, n_\mathrm{u}),\\
\mathcal{R}_1 &= \mathcal{A} *_{2,1} \mathcal{R}_0, && \dim(\mathcal{R}_1) = (n_\mathrm{x}, n_\mathrm{p}+1, n_\mathrm{p}+1, n_\mathrm{u}),\\
&\ \vdots && \qquad \vdots\\
\mathcal{R}_n &= \mathcal{A} *_{2,1} \mathcal{R}_{n-1}, && \dim(\mathcal{R}_n) = (n_\mathrm{x}, n_\mathrm{p}+1, \dots, n_\mathrm{p}+1, n_\mathrm{u}).
\end{aligned} \tag{31}$$

where $\tilde{\mathcal{B}}$ denotes the tensor $\mathcal{B}$ with the second and third dimension permuted, and $*_{2,1}$ denotes tensor-tensor product along the second dimension of tensor $\mathcal{A}$ and the first dimension of the tensor $\mathcal{R}_i$, for $i \in \mathbb{I}_1^\infty$. Note that the tensors are growing in dimension with $n$, and each horizon step adds a level of dependence on the scheduling variable.

Now, we can show that the evolution of the state can be represented with the reachability tensors. Let us begin with the initial state $x(0) = 0$, then the evolution of the state trajectory of $\Sigma$ can be rewritten as follows:

$$x(1) = B(p(0))u(0) = (B_0 + B_1 p_1(0) + \dots B_{n_\mathrm{p}} p_{n_\mathrm{p}}(0))u(0) = \mathcal{R}_0 \cdot_2 p^\top(0) \cdot_3 u^\top(0), \tag{32}$$

which shows that the state vector $x(1)$ lies in the image of the *reachability matrix* $R_0 = \begin{pmatrix} B_0 & B_1 & \dots & B_{n_\mathrm{p}} \end{pmatrix}$, but the same subspace can be obtained by the tensor singular vectors from the first dimension of reachability tensor $\mathcal{R}_0$. For the next time horizon:

$$\begin{aligned}
x(2) &= A(p(1))x(1) + B(p(1))u(1) =\\
&= (A_0 + A_1 p_1(1) + \dots A_{n_\mathrm{p}} p_{n_\mathrm{p}}(1))x(1) + (B_0 + B_1 p_1(1) + \dots B_{n_\mathrm{p}} p_{n_\mathrm{p}}(1))u(1) =\\
&= (\mathcal{A} * \mathcal{B}) \cdot_2 p^\top(1) \cdot_3 p^\top(0) \cdot_4 u^\top(0) + \mathcal{B} \cdot_2 p^\top(1) \cdot_3 u^\top(1) =\\
&= \mathcal{R}_1 \cdot_2 p^\top(1) \cdot_3 p^\top(0) \cdot_4 u^\top(0) + \mathcal{R}_0 \cdot_2 p^\top(1) \cdot_3 u^\top(1).
\end{aligned} \tag{33}$$

From the equations above, we can again see that $x(2) \in \mathrm{im}(R_1)$ for $R_1 = \begin{pmatrix} R_0 & A_0 R_0 & \dots & A_{n_\mathrm{p}} R_0 \end{pmatrix}$, and the same subspace can also be obtained by the collection of tensor singular vectors from the first dimension of reachability tensors $\mathcal{R}_1$ and $\mathcal{R}_0$.

Then, the state evolution for any $n \in \mathbb{N}$ can be represented with reachability tensors $\mathcal{R}_n, \dots, \mathcal{R}_0$ operating on the scheduling and input signals:

$$x(n+1) = A(p(n))x(n) + B(p(n))u(n) = \mathcal{R}_n \cdot_2 p^\top(n) \cdot_3 \dots \cdot_{n+2} p^\top(0) \cdot_{n+3} u^\top(0) + \dots + \mathcal{R}_0 \cdot_2 p^\top(n) \cdot_3 u^\top(n). \tag{34}$$

The original definition from[7] is here extended by showing that the state-reachable subspaces can be obtained from the tensor singular value decomposition, since the first dimension of reachability tensors is related to the states.

**Definition 17.** (Extension of $n$–partial state-reachability space of LPV-SS representations)

$$\begin{aligned} \mathfrak{R}_0(\Sigma) &= \operatorname{span} \bigcup_{j \in \mathbb{I}_0^{n_\mathrm{p}}} \operatorname{im}(B_j) = \operatorname{span}\{\phi_1(\mathcal{R}_0)\} = \operatorname{im}(R_0), \\ \mathfrak{R}_i(\Sigma) &= \mathfrak{R}_{i-1}(\Sigma) \cup \bigcup_{j \in \mathbb{I}_0^{n_\mathrm{p}}} \operatorname{im}(A_j R_{i-1}) = \mathfrak{R}_{i-1}(\Sigma) \cup \operatorname{span}\{\phi_1(\mathcal{R}_i)\} = \operatorname{im}(R_i), \end{aligned} \tag{35}$$

for $i \in \mathbb{I}_1^n$, where $A_j$, and $B_j$, for $j \in \mathbb{I}_0^{n_\mathrm{p}}$ are matrices from LPV-SS representation (19), $\phi_1(\mathcal{R}_i)$ represents the singular vectors of tensor $\mathcal{R}_i$ along the first dimension, and $R_i \in \mathbb{R}^{n_\mathrm{x} \times r_i}$ represents the reachability matrix of LPV-SS representation $\Sigma$ over horizon $i$.

More importantly, reachability tensors also show what are the dominant directions of scheduling variables affecting a state evolution. In the following example, it is shown how the singular vectors form a scheduling-reachable space (and similarly, a state-reachable space as well):

$$\begin{aligned} p(0) = \mathcal{R}_0 \cdot_1 x(1) \cdot_3 u(0) &= \Big(\sum_{i=1}^{R} \sigma_i \, \phi_1^{(i)} \otimes \phi_2^{(i)} \otimes \phi_3^{(i)}\Big) \cdot_1 x^\top(1) \cdot_3 u^\top(0) = \\ &= \sum_{i=1}^{R} \sigma_i \Big( (\phi_1^{(i)} \otimes \phi_2^{(i)} \otimes \phi_3^{(i)}) \cdot_1 x^\top(1) \cdot_3 u^\top(0) \Big) = \\ &= \sum_{i=1}^{R} \sigma_i \underbrace{(\phi_1^{(i)})^\top x(1)}_{\in \mathbb{R}} \otimes \phi_2^{(i)} \otimes \underbrace{(\phi_3^{(i)})^\top u(0)}_{\in \mathbb{R}} = \sigma_1 \phi_2^{(1)} + \dots + \sigma_R \phi_2^{(R)}. \end{aligned} \tag{36}$$

Similarly, the values of scheduling signals in the next time steps can be obtained as:

$$\begin{aligned} p(1) &= \mathcal{R}_1 \cdot_1 x^\top(2) \cdot_3 p^\top(0) \cdot_4 u^\top(0), \\ &\vdots \\ p(n) &= \mathcal{R}_n \cdot_1 x^\top(n+1) \cdot_2 p^\top(n-1) \cdots \cdot_{n+3} u^\top(0). \end{aligned}$$

Based on this, we can make the following definition:

**Definition 18.** ($n$–partial scheduling-reachability space of LPV-SS representations)

$$\begin{aligned} \mathcal{P}_0(\Sigma) &= \operatorname{span}\{\phi_2(\mathcal{R}_0)\} = \operatorname{im}(P_0), \\ \mathcal{P}_i(\Sigma) &= \mathcal{P}_{i-1}(\Sigma) \cup \operatorname{span}\{\phi_2(\mathcal{R}_i), \dots, \phi_{i+2}(\mathcal{R}_i)\} = \operatorname{im}(P_i), \end{aligned} \tag{37}$$

for $i \in \mathbb{I}_1^n$, where $\phi_2(\mathcal{R}_i)$ represents the singular vectors of the tensor $\mathcal{R}_i$ along the second dimension, and $P_i \in \mathbb{R}^{n_\mathrm{p} \times r_{p_i}}$ represents the scheduling-reachability matrix of LPV-SS representation $\Sigma$ over horizon $i$.

The elements from the the scheduling-reachability space $\mathcal{P}_n(\Sigma)$ are the only part of the scheduling variables that influence the states in the state-evolution equations. Thus, for the first two steps of the horizon ($n = 0$ and $n = 1$), we obtain:

$$\begin{aligned} B(p(n))u(n) &= B(\hat{p}(n))u(n), & \forall p(n) \in \mathbb{P},\ \forall u(n) \in \mathbb{U}, \\ A(p(n))B(p(n-1))u(n-1) &= A(\hat{p}(n))B(\hat{p}(n-1))u(n-1), & \forall p(n), p(n-1) \in \mathbb{P},\ \forall u(n-1) \in \mathbb{U}, \end{aligned} \tag{38}$$

where $\hat{p}$ refers to the projection of the scheduling variable $p$ onto the subspace $\mathcal{P}_N(\Sigma)$. This pattern continues for higher values of $N$.

Next, we define partial observability tensors of an affine LPV-SS model $\Sigma$ in a similar manner. Let $\mathcal{A}$ and $\mathcal{C}$ be the state and output tensors from the tensor form $\Phi$ of an LPV-SS model (21). Then, the observability tensors for horizon $n$ are defined recursively as:

$$
\begin{aligned}
\mathcal{O}_0 &= \tilde{\mathcal{C}}, & \dim(\mathcal{O}_0) &= (n_\mathrm{y}, n_\mathrm{p}+1, n_\mathrm{x}),\\
\mathcal{O}_1 &= \mathcal{O}_0 *_{3,1} \mathcal{A}, & \dim(\mathcal{O}_1) &= (n_\mathrm{y}, n_\mathrm{p}+1, n_\mathrm{p}+1, n_\mathrm{x}),\\
&\vdots & &\vdots\\
\mathcal{O}_n &= \mathcal{O}_{n-1} *_{n+2,1} \mathcal{A}, & \dim(\mathcal{O}_n) &= (n_\mathrm{y}, n_\mathrm{p}+1, \ldots, n_\mathrm{p}+1, n_\mathrm{x}).
\end{aligned} \tag{39}
$$

where $*$ denotes the tensor-tensor product along the specified dimensions of $\mathcal{O}_i$ and tensor $\mathcal{A}$, for $i \in \mathbb{I}_1^\infty$. Observability tensors are growing in order, since each new horizon step adds a level of dependance on the scheduling variable.

Now, we want to show that the output trajectories of LPS-SS representation $\Sigma$ can be represented with observability tensors. Let the input signal be $u \equiv 0$, then, the evolution of the output trajectory can be rewritten as:

$$
\begin{aligned}
y(0) &= C(p(0))x(0) = \mathcal{C} \cdot_2 x^\top(0) \cdot_3 p^\top(0) = \mathcal{O}_0 \cdot_2 p^\top(0) \cdot_3 x^\top(0),\\
y(1) &= C(p(1))A(p(0))x(0) = (\mathcal{C} *_{2,1} \mathcal{A}) \cdot_2 p^\top(1) \cdot_3 p^\top(0) \cdot_4 x^\top(0) = \mathcal{O}_1 \cdot_2 p^\top(1) \cdot_3 p^\top(0) \cdot_4 x^\top(0),\\
&\vdots\\
y(n) &= C(p(n))A(p(n-1)) \cdots A(p(0))x(0) = \mathcal{O}_n \cdot_2 p^\top(n) \cdot_3 \cdots \cdot_{n+2} p^\top(0) \cdot_{n+3} x^\top(0).
\end{aligned} \tag{40}
$$

Based on this analysis, we define the $N$-partial state-observability and scheduling-observability spaces for LPV-SS representations as follows:

**Definition 19.** (Extension of $n$– partial state-observability space of LPV-SS representations)

$$
\begin{aligned}
\mathcal{O}_0(\Sigma) &= \Big(\ker \bigcap_{j\in\mathbb{I}_0^{n_\mathrm{p}}} \mathrm{im}(C_j)\Big)^\perp = \mathrm{span}\{\phi_3(\mathcal{O}_0)\} = \mathrm{im}(O_0),\\
\mathcal{O}_i(\Sigma) &= \mathcal{O}_{i-1}(\Sigma) \cup \Big(\bigcap_{j\in\mathbb{I}_0^{n_\mathrm{p}}} \ker(O_{i-1}^\top A_j)\Big)^\perp = \mathcal{O}_{i-1}(\Sigma) \cup \mathrm{span}\{\phi_{i+3}(\mathcal{O}_i)\} = \mathrm{im}(O_i),
\end{aligned} \tag{41}
$$

for $i \in \mathbb{I}_1^n$, where $A_j$, and $C_j$ for $j \in \mathbb{I}_0^{n_\mathrm{p}}$ are matrices from the LPV-SS representation $\Sigma$ and each $\phi_j(\mathcal{O}_i)$ represents the singular vectors of the tensor $\mathcal{O}_i$ along dimension $j$, where $j \in \mathbb{I}_1^{i+3}$. Matrices $O_i$ are observability matrices of the LPV-SS representation $\Sigma$ at horizon $i$.

**Definition 20.** ($n$–partial observability space of scheduling variables of LPV-SS representations)

$$
\begin{aligned}
\mathcal{Q}_0(\Sigma) &= \mathrm{span}\{\phi_2(\mathcal{O}_0)\} = \mathrm{im}(Q_0),\\
\mathcal{Q}_i(\Sigma) &= \mathcal{Q}_{i-1}(\Sigma) \cup \mathrm{span}\{\phi_2(\mathcal{O}_i), \ldots, \phi_{i+2}(\mathcal{O}_i)\} = \mathrm{im}(Q_i),
\end{aligned} \tag{42}
$$

for $i \in \mathbb{I}_1^N$, where $\phi_2(\mathcal{O}_i)$ represents the singular vectors of the tensor $\mathcal{O}_i$ along the second dimension and $Q_i \in \mathbb{R}^{n_\mathrm{p} \times r_{p_i}}$ represents the scheduling-observability matrix of LPV-SS representation $\Sigma$ over horizon $i$.

Similarly as for the reachability case, the elements from the the scheduling-observability space $\mathcal{Q}_n(\Sigma)$ are the only part of the scheduling variables that influence the observability of the states from the outputs. Thus, for the first two steps of the horizon ($n = 0$ and $n = 1$), we obtain:

$$
\begin{aligned}
C(p(n))x(n) &= C(\hat{p}(n))x(n), & &\forall p(n) \in \mathbb{P},\ \forall x(n) \in \mathbb{X},\\
C(p(n))A(p(n-1))x(n-1) &= C(\hat{p}(n))A(\hat{p}(n-1))x(n-1) & &\forall p(n), p(n-1) \in \mathbb{P},\ \forall x(n-1) \in \mathbb{X},
\end{aligned} \tag{43}
$$

where $\hat{p}$ refers to the projection of the scheduling variable $p$ onto the subspace $\mathcal{Q}_N(\Sigma)$. This pattern continues for higher values of $n$.

**TABLE 1** LPV Petrov-Galerkin projection matrices for tensor-based LPV moment matching method

| | Reachability mode | Observability mode | Hankel mode |
|---|---|---|---|
| State projection | $V = R_n \in \mathbb{R}^{n_x \times r_x}$ | $V = O_n \in \mathbb{R}^{n_x \times r_x}$ | $V = R_n R^{-1} \in \mathbb{R}^{n_x \times r_x}$ |
| Residual projection | $W = R_n \in \mathbb{R}^{n_x \times r_x}$ | $W = O_n \in \mathbb{R}^{n_x \times r_x}$ | $W = O^{-1} O_n^\top \in \mathbb{R}^{n_x \times r_x}$ |
| Scheduling projection | $Z = P_n \in \mathbb{R}^{(n_p+1) \times r_p}$ | $Z = Q_n \in \mathbb{R}^{(n_p+1) \times r_p}$ | $Z = \text{orth}\begin{pmatrix} P_n & Q_n \end{pmatrix} \in \mathbb{R}^{(n_p+1) \times r_p}$ |

So far, we have defined the reachability and observability tensors and demonstrated how they provide insight into the essential state and scheduling subspaces. The matrices that characterize these subspaces can be used to construct the projection matrices for the extended LPV Petrov-Galerkin approach from Section 4 to derive the reduced model.

We introduce three reduction modes, as summarized in Table 1, each ensuring that a specific aspect of the Markov parameters remains equivalent between the original and the reduced model:

1. **Reachability-based reduction**: This mode constructs reduced-order state and scheduling spaces based on the reachability tensors up to horizon $n$. It ensures that the entires in the reachability form:

$$B(p(t)), \quad A(p(t))B(p(t-1)), \quad A(p(t))A(p(t-1))B(p(t-2)), \quad \ldots$$

   for the first $n$ steps are exactly preserved in the reduced model. That is, for all $k \leq n$,

$$W^\top A_r(p_r(t)) \ldots A_r(p_r(t-k-1))B_r(p_r(t-k)) = A(p(t)) \cdots A(p(t-k-1))B(p(t-k)).$$

2. **Observability-based reduction**: This mode constructs reduced-order state and scheduling spaces based on the observability tensors up to horizon $N$. It ensures that the entires in the observability form:

$$C(p(t)), \quad C(p(t+1))A(p(t)), \quad C(p(t+2))A(p(t+1))A(p(t)), \quad \ldots$$

   for the first $n$ steps are exactly preserved in the reduced model. That is, for all $k \leq n$,

$$C_r(p_r(t+k))A_r(p_r(t+k-1)) \ldots A_r(p_r(t))V = C(p(t+k))A(p(t+k-1)) \cdots A(p(t)).$$

3. **Hankel-based reduction**: This mode aims at the preservation of the Markov parameters which are combinations of the entries of the reachability and observability tensors:

$$C(p(t))B(p(t-1)), \quad C(p(t))A(p(t-1))B(p(t-2)), \quad \ldots$$

   More precisely, the input-output trajectories are determined by the state-evolution (34) and the output-evolution (40), allowing to characterize the next $N$ outputs $y(t+1), \ldots, y(t+n)$ in terms of the $n$ inputs $u(t), \ldots u(t+n-1)$. The relation can also be written using Kroneker products and extended k-step reachability and observability matrices as in[26], however, the tensor approach adds the scheduling projection as well. The corresponding reduction effectively minimizes both $\dim_x(\Sigma_r)$ and $\dim_p(\Sigma_r)$ while preserving the most important dynamical information over a horizon $2n$.

Each of the discussed modes ensures equivalence of the corresponding aspects between the FOM and ROM, guaranteeing consistency between the models up to the chosen horizon.

Algorithm 2 shows the computation of a matrix representations of state-reachability space and scheduling-reachability space for a given horizon $n$, resulting in matrices $R_n \in \mathbb{R}^{n_x \times r_x}$ and $P_n \in \mathbb{R}^{n_p \times r_p}$. Algorithm 3 shows the same construction for the state-observability space and scheduling-observability space for a given horizon $n$, resulting in matrices $O_n \in \mathbb{R}^{n_x \times r_x}$ and $Q_n \in \mathbb{R}^{n_p \times r_p}$. Finally, the full tensor-based moment matching for LPV-SS representation is given in Algorithm 4.

The process of obtaining tensor singular vectors is inherently complex, as different decompositions, such as TSVD and HOSVD, yield varying results with different complexities and accuracy levels. While TSVD offers an exact solution for diagonalizable tensors, it is computationally more expensive than HOSVD. For non-diagonalizable tensors, both methods approximate the dominant directions, which can introduce inaccuracies in the reduced model. The induced error depends on the decomposition method and the properties of the tensor with trade-offs between computational complexity and model accuracy. As a result, the final reduced model may contain errors due to the limitations of the chosen decomposition approach.

## 5.2 The POD Method for LPV-SS models

The POD method provides an optimally ordered, orthonormal basis in the least-squares sense for a given set of simulation or experimental data from the system. It is commonly used with Petrov-Galerkin projection to approximate complex systems. In general, POD problem can be formulated as follows:

Let $w_1, \dots, w_N \in \mathcal{W}$ be a set of data vectors in a Hilbert space $\mathcal{W}$. The POD basis is defined as the orthonormal set $\{\varphi_k\}_{k=1}^{\infty} \subset \mathcal{W}$ that minimizes the average projection error for any truncation level $r$:

$$J(\varphi_1, \dots, \varphi_r) = \sum_{j=1}^{N} \left\| w_j - \sum_{k=1}^{r} \langle w_j, \varphi_k \rangle \varphi_k \right\|_2^2 . \tag{44}$$

The optimal basis vectors $\varphi_k$ correspond to the left singular vectors of the data matrix, and the associated singular values quantify the approximation quality.

We now aim to formulate how the POD method can be applied to LPV-SS models. In this setting, simulation data is collected from the system in the form of state and scheduling trajectories, typically generated to explore the relevant operating region of the system. The objective is to identify low-dimensional subspaces for both the state and scheduling domains, which can then be used to construct a reduced-order model through the LPV Petrov Galerking projection from Section 4.

To apply dimensionality reduction to LPV-SS models, we begin by considering low-rank approximations of the simulation trajectories of both the state $x \in \mathcal{X}$ and the scheduling signal $p \in \mathcal{P}$.

**Algorithm 2** Calculate matrix representations of state-reachability space $\mathcal{R}_n(\Sigma)$ and scheduling-reachability space $\mathcal{P}_n(\Sigma)$

**Input:** tensors $(\mathcal{A}, \mathcal{B})$ of LPV-SS model $\Sigma$, horizon $n \in \mathbb{N}$

**Output:** $R_n \in \mathbb{R}^{n_x \times r_x}$ and $P_n \in \mathbb{R}^{(n_p+1) \times r_p}$ (such that $\mathrm{im}(R_n) = \mathcal{R}_n(\Sigma)$ and $\mathrm{im}(P_n) = \mathcal{P}_n(\Sigma)$)

`construct:` initial reachability tensor $\mathcal{R}_0$ as in (31)
`apply:` tensor decomposition on $\mathcal{R}_0$
`update:` $R_n \leftarrow \phi_1(\mathcal{R}_0)$ and $P_n \leftarrow \phi_2(\mathcal{R}_0)$
`for k = 1:n do`
  `construct:` reachability tensor $\mathcal{R}_k$ via (31)
  `apply:` tensor decomposition on $\mathcal{R}_k$
  `update:` $R_n \leftarrow \mathrm{orth}(R_n, \phi_1(\mathcal{R}_k))$ and $P_n \leftarrow \mathrm{orth}(P_n, \phi_2(\mathcal{R}_k), \dots, \phi_{k+2}(\mathcal{R}_k))$
`end for`
`return:` $R_n, P_n$

**Algorithm 3** Calculate matrix representations of state-observability space $\mathcal{O}_n(\Sigma)$ and scheduling-observability space $\mathcal{Q}_n(\Sigma)$

**Input:** tensors $(\mathcal{A}, \mathcal{C})$ of LPV-SS model $\Sigma$, horizon $n \in \mathbb{N}$

**Output:** $O_n \in \mathbb{R}^{n_x \times r_x}$ and $Q_n \in \mathbb{R}^{(n_p+1) \times r_p}$ (such that $\mathrm{im}(O_n) = \mathcal{O}_n(\Sigma)$ and $\mathrm{im}(Q_n) = \mathcal{Q}_n(\Sigma)$)

`construct:` initial observability tensor $\mathcal{O}_0$ as in (39)
`apply:` tensor decomposition on $\mathcal{O}_0$
`update:` $O_n \leftarrow \phi_2(\mathcal{O}_0)$ and $Q_n \leftarrow \phi_3(\mathcal{O}_0)$
`for k = 1:n do`
  `construct:` observability tensor $\mathcal{O}_k$ via (39)
  `apply:` tensor decomposition on $\mathcal{O}_k$
  `update:` $O_n \leftarrow \mathrm{orth}(O_n, \phi_{k+1}(\mathcal{O}_k))$ and $Q_n \leftarrow \mathrm{orth}(Q_n, \phi_2(\mathcal{O}_k), \dots, \phi_{k+2}(\mathcal{O}_k))$
`end for`
`return:` $O_n, Q_n$

**Algorithm 4** Tensor-based moment matching for LPV-SS representations

**Inputs:** $\Sigma = (\mathcal{A}, \mathcal{B}, \mathcal{C}, \mathcal{D}, \eta)$, mode $\in$ {R,O,H}, decomp $\in$ {HOSVD, TSVD}, horizon $n \in \mathbb{N}$
**Output:** $\Sigma_r = (\mathcal{A}_r, \mathcal{B}_r, \mathcal{C}_r, \mathcal{D}_r, \eta_r)$

**call:** Algorithms 2 and 3 to compute $R_n, P_n, O_n, Q_n$
**if** Mode = 'R' **then**
  **update:** $V \leftarrow R_n$, $W \leftarrow R_n$, $Z \leftarrow P_n$
**end if**
**if** Mode = 'O' **then**
  **update:** $V \leftarrow O_n$, $W \leftarrow O_n$, $Z \leftarrow Q_n$
**end if**
**if** Mode = 'T' and $\mathrm{rank}(R_n) = \mathrm{rank}(O_n) = \mathrm{rank}(O_n^\top R_n)$ **then**
  **calculate:** $(R, O) \leftarrow$ factorization of $O_n^\top R_n$
  **update:** $V \leftarrow R_n R^{-1}$, $W \leftarrow O_n^\top O^{-1}$, $Z \leftarrow \mathrm{orth}\begin{pmatrix} P_n & Q_n \end{pmatrix}$
**end if**
**calculate:** $(\Sigma_r, \eta_r)$ using Algorithm 1 and V, W, and Z.
**return:** $\Sigma_r, \eta_r$

We thus consider spectral decompositions of the form:

$$x(t) = \sum_{k=1}^{n_\mathrm{x}} a_k^{(x)}(t)\, \varphi_k^{(x)}, \quad p(t) = \sum_{k=1}^{n_\mathrm{p}} a_k^{(p)}(t)\, \varphi_k^{(p)}, \tag{45}$$

where $\{\varphi_k^{(x)}\} \subset \mathbb{X}$ and $\{\varphi_k^{(p)}\} \subset \mathbb{P}$ are orthonormal basis vectors derived from the collected data. Since the signals live in finite-dimensional spaces $\mathbb{X} \cong \mathbb{R}^{n_\mathrm{x}}$ and $\mathbb{P} \cong \mathbb{R}^{n_\mathrm{p}}$, each snapshot can be expressed using a finite number of basis functions. The approximation of state vector $\hat{x}(t) \in \mathcal{V} \subset \mathbb{X}$ and the approximation of scheduling vector $\hat{p}(t) \in \mathcal{Z} \subset \mathbb{P}$ are obtained by truncating the expansions to $r_\mathrm{x}$ and $r_\mathrm{p}$ terms, respectively.

Truncating these series yields lower-rank approximations of the original signals:

$$\hat{x}(t) = \sum_{k=1}^{r_\mathrm{x}} a_k^{(x)}(t)\, \varphi_k^{(x)}, \quad \hat{p}(t) = \sum_{k=1}^{r_\mathrm{p}} a_k^{(p)}(t)\, \varphi_k^{(p)}. \tag{46}$$

At this stage, reduced bases are typically chosen to minimize signal reconstruction error, leading to the independent computation of projection matrices where state and scheduling subspaces are computed separately. However, in model reduction, the goal is to preserve the input-output behavior of the system, and minimizing signal error does not guarantee preservation of this behavior.

When state and scheduling trajectories interact strongly, as is often the case in LPV systems, independent reductions may fail to capture important coupled dynamics. This motivates the joint computation of projection matrices, where both state and scheduling data are considered together. Structured data tensors are used to enable coupled dimensionality reduction, preserving system dynamics more effectively.

### 5.2.1 | Independent Computation of Projection Matrices

Assume we have collected the following data sets containing state, scheduling, and the state increments from simulations of a persistently excited LPV system:

$$\mathcal{D}_x = \{x(k)\}_{k=0}^{N-1}, \quad \mathcal{D}_p = \{p(k)\}_{k=0}^{N-1}, \quad \mathcal{D}_\Delta = \{\Delta_x(k)\}_{k=0}^{N-1}, \tag{47}$$

where $\Delta_x(k) = x(k+1) - x(k),\ k \in \mathbb{I}_0^{N-2}$. Let the collected data be stored in the snapshot matrices:

$$M_\mathrm{x} = \begin{pmatrix} x(0) & x(1) & \cdots & x(N-1) \end{pmatrix} \in \mathbb{R}^{n_\mathrm{x} \times N},$$
$$M_\mathrm{p} = \begin{pmatrix} p(0) & p(1) & \cdots & p(N-1) \end{pmatrix} \in \mathbb{R}^{n_\mathrm{p} \times N}, \tag{48}$$

with $\rho_\mathrm{x} = \mathrm{rank}(M_\mathrm{x}), \rho_\mathrm{p} = \mathrm{rank}(M_\mathrm{p})$, and sets of singular vectors $\{\varphi_k^{(x)}\}_{k=1}^{\rho_\mathrm{x}}$, $\{\varphi_k^{(p)}\}_{k=1}^{\rho_\mathrm{p}}$, and singular values $\{\sigma_k^{(x)}\}_{k=1}^{\rho_\mathrm{x}}$, $\{\sigma_k^{(p)}\}_{k=1}^{\rho_\mathrm{p}}$ of $M_\mathrm{x}, M_\mathrm{p}$, respectively. The matrices $M_\mathrm{x}$ and $M_\mathrm{p}$ contain snapshots from the state and scheduling spaces $\mathbb{X}$ and $\mathbb{P}$, respectively. Since the singular vectors of $M_\mathrm{x}$ form an orthonormal basis for the column space of the data, any state vector $x(t)$ can be expanded using these vectors. Their projection on a lower-dimensional subspace is equal to the truncated sum:

$$x(t) = \sum_{i=1}^{n_\mathrm{x}} \langle x(t), \varphi_i^{(x)} \rangle \varphi_i^{(x)}, \quad \hat{x}(t) = \sum_{i=1}^{r_\mathrm{x}} \langle x(t), \varphi_i^{(x)} \rangle \varphi_i^{(x)}, \tag{49}$$

the error of state projection on the simulation data $\mathcal{D}_x$ for any orthonormal basis can be written as:

$$J_\mathrm{x}(\varphi_1^{(x)}, \ldots, \varphi_{r_\mathrm{x}}^{(x)}) = \sum_{j=0}^{N-1} \|x(j) - \hat{x}(j)\|^2 = \sum_{k=r_\mathrm{x}+1}^{n_\mathrm{x}} (\varphi_k^{(x)})^\top M_\mathrm{x} M_\mathrm{x}^\top (\varphi_k^{(x)}).$$

For a detailed proof see the Appendix. The cost function $J_\mathrm{x}(\varphi_1^{(x)}, \ldots, \varphi_{r_\mathrm{x}}^{(x)})$ is minimal for every truncation level $r_\mathrm{x} = 1, \ldots, n_\mathrm{x}$, if and only if the basis vectors $\varphi_1^{(x)}, \ldots, \varphi_{r_\mathrm{x}}^{(x)}$ are the first $r_\mathrm{x}$ singular vectors of the data-matrix with states $M_\mathrm{x}$. Then, the approximation error of optimal state projection is:

$$J_\mathrm{x}^*(\varphi_1^{(x)}, \ldots, \varphi_{r_\mathrm{x}}^{(x)}) = \sum_{k=r_\mathrm{x}+1}^{n_\mathrm{x}} (\sigma_k^{(x)})^2. \tag{50}$$

Similarly, the error of scheduling projection on the simulation data $\mathcal{D}_p$ for any orthonormal basis is equal to

$$J_\mathrm{p}(\varphi_1^{(p)}, \ldots, \varphi_{r_\mathrm{p}}^{(p)}) = \sum_{k=r_\mathrm{p}+1}^{n_\mathrm{p}} (\varphi_k^{(p)})^\top M_\mathrm{p} M_\mathrm{p}^\top (\varphi_k^{(p)}), \tag{51}$$

and is minimal for every truncation level $r_\mathrm{p} = 1, \ldots, n_\mathrm{p}$, if and only if the basis vectors $\varphi_1^{(p)}, \ldots, \varphi_{r_\mathrm{p}}^{(p)}$ are the first $r_\mathrm{p}$ singular vectors of the data-matrix with scheduling variables $M_\mathrm{p}$. Then, the approximation error of the optimal scheduling projection is:

$$J_\mathrm{p}^*(\varphi_1^{(p)}, \ldots, \varphi_{r_\mathrm{p}}^{(p)}) = \sum_{k=r_\mathrm{p}+1}^{n_\mathrm{p}} (\sigma_k^{(p)})^2. \tag{52}$$

We can then obtain the projection matrices that can be used in the Petrov-Galerkin projection proposed in Section 4 to obtain the reduced model $\Sigma_r$ of the form (25), where

$$V = \begin{pmatrix} \varphi_1^{(x)} & \cdots & \varphi_{r_\mathrm{x}}^{(x)} \end{pmatrix} \in \mathbb{R}^{n_\mathrm{x} \times r_\mathrm{x}}, \quad Z = \begin{pmatrix} \varphi_1^{(p)} & \cdots & \varphi_{r_\mathrm{p}}^{(p)} \end{pmatrix} \in \mathbb{R}^{n_\mathrm{p} \times r_\mathrm{p}}, \quad W = \begin{pmatrix} \varphi_1^{(\Delta_x)} & \cdots & \varphi_{r_\mathrm{x}}^{(\Delta_x)} \end{pmatrix}, \quad \text{or } W = V.$$

### 5.2.2 | Joint Computation of Projection Matrices

We now focus on computing the projection matrices jointly. Simulation data is gathered as explained in (47), but instead of organizing it into matrices, we define a tensor $\mathcal{M} \in \mathbb{R}^{n_\mathrm{x} \times n_\mathrm{p} \times N}$ by stacking the outer products of the state and scheduling trajectories across the third dimension:

$$\mathcal{M} = \sum_{t=1}^{N} x(t) \otimes p(t) \otimes e_t, \tag{53}$$

where $e_t$ is the $t$-th unit vector in $\mathbb{R}^N$. The tensor $\mathcal{M}$ captures the structure of the data in the product space $\mathbb{X} \times \mathbb{P}$. In this setting, the approximation is performed jointly on the reduced product space $\mathbb{X}_\mathrm{r} \times \mathbb{P}_\mathrm{r}$, rather than independently on $\mathbb{X}$ and $\mathbb{P}$, in order to capture interactions between the state and scheduling variables more effectively.

The cost function (54) is formulated based on the data and it represents the projection error of the elements from the product space $\mathbb{X} \times \mathbb{P}$. Let $\phi^{(x)}$ denote the set of basis vectors for state variables and $\phi^{(p)}$ denote the set of basis vectors for scheduling variables. The product space is equipped with a tensor inner product $\langle \cdot, \cdot \rangle$ and the corresponding induced norm $\| \cdot \|_\circ$. Then,

$$J_\mathrm{xp}(\phi^{(x)}, \phi^{(p)}) = \sum_{t=0}^{N-1} \|x(t) \otimes p(t) - \hat{x}(t) \otimes \hat{p}(t)\|_\circ^2 = \sum_{t=0}^{N-1} \left\| x(t)p^\top(t) - \hat{x}(t)\hat{p}^\top(t) \right\|_\circ^2 \tag{54}$$

which can be written as

$$\begin{aligned} J_\mathrm{xp}(\phi^{(x)}, \phi^{(p)}) &= \sum_{t=0}^{N-1} \left\| \Big( \sum_{i=1}^{n_\mathrm{x}} \langle x(t), \phi_i^{(x)} \rangle \phi_i^{(x)} \Big) \Big( \sum_{i=1}^{n_\mathrm{p}} \langle p(t), \phi_i^{(p)} \rangle \phi_i^{(p)} \Big)^\top - \Big( \sum_{i=1}^{r_\mathrm{x}} \langle x(t), \phi_i^{(x)} \rangle \phi_i^{(x)} \Big) \Big( \sum_{i=1}^{r_\mathrm{p}} \langle p(t), \phi_i^{(p)} \rangle \phi_i^{(p)} \Big)^\top \right\|_\circ^2 = \\ &= \sum_{t=0}^{N-1} \left\| \Big( \sum_{i=1}^{n_\mathrm{x}} \sum_{j=1}^{n_\mathrm{p}} \langle x(t), \phi_i^{(x)} \rangle \langle p(t), \phi_j^{(p)} \rangle \phi_i^{(x)} \otimes \phi_j^{(p)} \Big) - \Big( \sum_{i=1}^{r_\mathrm{x}} \sum_{j=1}^{r_\mathrm{p}} \langle x(t), \phi_i^{(x)} \rangle \langle p(t), \phi_j^{(p)} \rangle \phi_i^{(x)} \otimes \phi_j^{(p)} \Big) \right\|_\circ^2 = \\ &= \sum_{t=0}^{N-1} \left\| \Big( \sum_{i=r_\mathrm{x}+1}^{n_\mathrm{x}} \sum_{j=1}^{r_\mathrm{p}} \langle x(t), \phi_i^{(x)} \rangle \langle p(t), \phi_j^{(p)} \rangle \phi_i^{(x)} \otimes \phi_j^{(p)} \Big) + \Big( \sum_{i=1}^{n_\mathrm{x}} \sum_{j=r_\mathrm{p}+1}^{n_\mathrm{p}} \langle x(t), \phi_i^{(x)} \rangle \langle p(t), \phi_j^{(p)} \rangle \phi_i^{(x)} \otimes \phi_j^{(p)} \Big) \right\|_\circ^2 \end{aligned}$$

Now, by defining $\alpha_{i,j} = \langle x(t), \phi_i^{(x)} \rangle \langle p(t), \phi_j^{(p)} \rangle \phi_i^{(x)} \otimes \phi_j^{(p)}$, we obtain:

$$J_\mathrm{xp}(\phi^{(x)}, \phi^{(p)}) = \sum_{t=0}^{N-1} \left\langle \sum_{i=r_\mathrm{x}+1}^{n_\mathrm{x}} \sum_{j=1}^{r_\mathrm{p}} \alpha_{i,j} + \sum_{i=1}^{n_\mathrm{x}} \sum_{j=r_\mathrm{p}+1}^{n_\mathrm{p}} \alpha_{i,j}, \sum_{k=r_\mathrm{x}+1}^{n_\mathrm{x}} \sum_{l=1}^{r_\mathrm{p}} \alpha_{k,l} + \sum_{k=1}^{n_\mathrm{x}} \sum_{l=r_\mathrm{p}+1}^{n_\mathrm{p}} \alpha_{k,l} \right\rangle. \tag{55}$$

Let us take further analyze the tensor inner-product term in the Equation (55). For orthonormal basis vectors, the following relation holds:

$$\left\langle \phi_i^{(x)} \otimes \phi_j^{(p)}, \phi_k^{(x)} \otimes \phi_l^{(p)} \right\rangle = \begin{cases} 1, & \text{for} \quad i = k \quad \text{and} \quad j = l, \\ 0, & \text{for} \quad i \neq k \quad \text{or} \quad j \neq l, \end{cases} \tag{56}$$

which implies that

$$\left\langle \alpha_{i,j}, \alpha_{k,l} \right\rangle = \begin{cases} \langle x(t), \phi_i^{(x)} \rangle^2 \cdot \langle p(t), \phi_j^{(p)} \rangle^2, & \text{for} \quad i = k \quad \text{and} \quad j = l, \\ 0, & \text{for} \quad i \neq k \quad \text{or} \quad j \neq l, \end{cases} \tag{57}$$

By exploiting (57) for $J_\mathrm{xp}(\phi^{(x)}, \phi^{(p)})$ and using the property of linearity of an inner product, we get

$$J_\mathrm{xp}(\phi^{(x)}, \phi^{(p)}) = \sum_{t=0}^{N-1} \left( \sum_{i=r_\mathrm{x}+1}^{n_\mathrm{x}} \sum_{j=1}^{r_\mathrm{p}} \langle x(t), \phi_i^{(x)} \rangle^2 \langle p(t), \phi_j^{(p)} \rangle^2 + \sum_{i=1}^{n_\mathrm{x}} \sum_{j=r_\mathrm{p}+1}^{n_\mathrm{p}} \langle x(t), \phi_i^{(x)} \rangle^2 \langle p(t), \phi_j^{(p)} \rangle^2 \right). \tag{58}$$

To preserve the symmetry in the expression, $J_\mathrm{xp}$ can be decomposed into three non-negative components, capturing different mismatch effects due to separate projections in the state and scheduling domains. Specifically,

$$J_\mathrm{xp}(\phi^{(x)}, \phi^{(p)}) = J_\mathrm{xp}^A(\phi^{(x)}, \phi^{(p)}) + J_\mathrm{xp}^B(\phi^{(x)}, \phi^{(p)}) - J_\mathrm{xp}^C(\phi^{(x)}, \phi^{(p)}) \tag{59}$$

with the individual terms defined as:

$$
\begin{aligned}
J^A_{\mathrm{xp}}(\phi^{(x)},\phi^{(p)}) &= \sum_{t=0}^{N-1}\sum_{i=r_\mathrm{x}+1}^{n_\mathrm{x}}\sum_{j=1}^{n_\mathrm{p}}\langle x(t),\phi_i^{(x)}\rangle^2\langle p(t),\phi_j^{(p)}\rangle^2,\\
J^B_{\mathrm{xp}}(\phi^{(x)},\phi^{(p)}) &= \sum_{t=0}^{N-1}\sum_{i=1}^{n_\mathrm{x}}\sum_{j=r_\mathrm{p}+1}^{n_\mathrm{p}}\langle x(t),\phi_i^{(x)}\rangle^2\langle p(t),\phi_j^{(p)}\rangle^2,\\
J^C_{\mathrm{xp}}(\phi^{(x)},\phi^{(p)}) &= \sum_{t=0}^{N-1}\sum_{i=r_\mathrm{x}+1}^{n_\mathrm{x}}\sum_{j=r_\mathrm{p}+1}^{n_\mathrm{p}}\langle x(t),\phi_i^{(x)}\rangle^2\langle p(t),\phi_j^{(p)}\rangle^2,
\end{aligned}
\tag{60}
$$

where each term quantifies the following:

- $J^A_{\mathrm{xp}}$: the loss due to projecting only the state while keeping the full scheduling space;
- $J^B_{\mathrm{xp}}$: the loss due to projecting only the scheduling variable while keeping the full state space;
- $J^C_{\mathrm{xp}}$: the residual loss due to joint projection, already accounted for by the overlap in $J^A_{\mathrm{xp}}$ and $J^B_{\mathrm{xp}}$.

Next, we focus on minimizing the individual terms, starting with $J^A_{\mathrm{xp}}$ to achieve the upper bound on the cost $J_{\mathrm{xp}}$.

$$
\begin{aligned}
J^A_{\mathrm{xp}}(\phi^{(x)},\phi^{(p)}) &= \sum_{t=0}^{N-1}\left(\sum_{i=r_\mathrm{x}+1}^{n_\mathrm{x}}\sum_{j=1}^{n_p}\langle x(t),\phi_i^{(x)}\rangle^2\langle p(t),\phi_j^{(p)}\rangle^2\right) = \\
&= \sum_{i=r_\mathrm{x}+1}^{n_\mathrm{x}}\sum_{t=0}^{N-1}\langle x(t),\phi_i^{(x)}\rangle^2\Big(\sum_{j=1}^{n_p}\langle p(t),\phi_j^{(p)}\rangle^2\Big).
\end{aligned}
$$

Since $\sum_{j=1}^{n_p}\langle p(t),\phi_j^{(p)}\rangle^2 = \|p(t)\|_2^2$ for any orthonormal basis $\{\phi_j^{(p)}\}_{j=1}^{n_p}$ $J_{\mathrm{xp}} = J_{\mathrm{xp}}(\phi^{(x)})$ is a function of only the **state** basis vectors.

$$
J^A_{\mathrm{xp}}(\phi^{(x)}) = \sum_{i=r_\mathrm{x}+1}^{n_\mathrm{x}}\sum_{t=0}^{N-1}\|p(t)\|_2^2(\phi_i^{(x)})^\top x(t)x^\top(t)\phi_i^{(x)} = \sum_{i=r_\mathrm{x}+1}^{n_\mathrm{x}}(\phi_i^{(x)})^\top\Big(\sum_{t=0}^{N-1}\|p(t)\|_2^2 x(t)x^\top(t)\Big)\phi_i^{(x)}
$$

Now, we can define a matrix $\widetilde{M}_\mathrm{x} = \sum_{t=0}^{N-1}\|p(t)\|_2^2 x(t)x^\top(t)$, which leads to the following expression:

$$
J^A_{\mathrm{xp}}(\phi^{(x)}) = \sum_{i=r_\mathrm{x}+1}^{n_\mathrm{x}}(\phi_i^{(x)})^\top\widetilde{M}_\mathrm{x}\phi_i^{(x)}. \tag{61}
$$

The minimal value of $J^A_{\mathrm{xp}}$ is achieved when $\phi_i^{(x)}$ are eigenvectors of $\widetilde{M}_\mathrm{x}$, corresponding to the smallest eigenvalues:

$$
\min_{\phi^{(x)}} J^A_{\mathrm{xp}} = \sum_{i=r_\mathrm{x}+1}^{n_\mathrm{x}}\lambda_i(\widetilde{M}_\mathrm{x}), \tag{62}
$$

where $\lambda_i(\widetilde{M}_\mathrm{x})$ are the eigenvalues of $\widetilde{M}_\mathrm{x}$ in decreasing order. Matrix $\widetilde{M}_\mathrm{x}$ is positive semi-definite, hence all eigenvectors and eigenvalues are guaranteed to be real and the eigenvalues are all non-negative and can be ordered in terms of their magnitude.

Next, we minimize $J^B_{\mathrm{xp}}$ to obtain the upper bound on $J_{\mathrm{xp}}$. The process is equivalent to the minimization of $J^A_{\mathrm{xp}}$:

$$
J^B_{\mathrm{xp}}(\phi^{(x)},\phi^{(p)}) = \sum_{t=0}^{N-1}\left(\sum_{j=r_\mathrm{p}+1}^{n_\mathrm{p}}\left\|x(t)\right\|^2\langle p(t),\phi_j^{(p)}\rangle^2\right) = \sum_{j=r_\mathrm{p}+1}^{n_\mathrm{p}}\sum_{t=0}^{N-1}\|x(t)\|_2^2(\phi_j^{(p)})^\top p(t)p^\top(t)\phi_j^{(p)} = \sum_{j=r_\mathrm{p}+1}^{n_\mathrm{p}}(\phi_j^{(p)})^\top\widetilde{M}_\mathrm{p}\phi_j^{(p)},
$$

where $\widetilde{M}_{\mathrm{p}} = \sum_{t=0}^{N-1} \|x(t)\|_2^2 p(t)p^\top(t)$. The minimal value is obtained when $\phi_j^{(p)}$ are eigenvectors of $\widetilde{M}_{\mathrm{p}}$:

$$\min_{\phi^{(p)}} J_{\mathrm{xp}}^B = \sum_{j=r_{\mathrm{p}}+1}^{n_{\mathrm{p}}} \lambda_j(\widetilde{M}_{\mathrm{p}}), \tag{63}$$

where $\lambda_j(\widetilde{M}_{\mathrm{p}})$ are the real non-negative eigenvalues in decreasing order.

**Theorem 2** (Upper Bound on Cross-Term Cost)**.** *For any orthonormal set of $\phi_x$ and $\phi_p$ defining the signal projections for $x(t)$ and $p(t)$, respectively, the following inequality holds:*

$$J_{\mathrm{xp}}(\phi^{(x)}, \phi^{(p)}) \leq J_{\mathrm{xp}}^A(\phi^{(x)}, \phi^{(p)}) + J_{\mathrm{xp}}^B(\phi^{(x)}, \phi^{(p)}), \tag{64}$$

*where the individual terms $J_{\mathrm{xp}}^A$, $J_{\mathrm{xp}}^B$, and $J_{\mathrm{xp}}^C$ are defined as in* (60)*.*

*Proof.* From the decomposition in (59) and the non-negativity of $J_{\mathrm{xp}}^C$, the inequality (64) follows directly. □

Next, we analyze the coupling error term $J_{\mathrm{xp}}^C$, which captures the interaction between the discarded components of the state and scheduling variables. It can be expressed as

$$J_{\mathrm{xp}}^C(\phi^{(x)}, \phi^{(p)}) = \sum_{t=0}^{N-1} \left( \sum_{i=r_{\mathrm{x}}+1}^{n_{\mathrm{x}}} \langle x(t), \phi_i^{(x)} \rangle^2 \right) \left( \sum_{j=r_{\mathrm{p}}+1}^{n_{\mathrm{p}}} \langle p(t), \phi_j^{(p)} \rangle^2 \right) = \sum_{t=0}^{N-1} \|\tilde{x}(t)\|_2^2 \|\tilde{p}(t)\|_2^2,$$

where $\tilde{x}(t)$ and $\tilde{p}(t)$ denote the discarded parts of the state and scheduling signals.

**Theorem 3** (Bounds on the coupling error term $J_{xp}^C$)**.** *For any orthonormal set of $\phi_x$ and $\phi_p$, the coupling error term $J_{xp}^C$ satisfies the following bounds:*

$$0 \leq J_{xp}^C \leq \min(J_{\mathrm{xp}}^A, J_{\mathrm{xp}}^B).$$

*Proof.* Term $J_{xp}^C$ is contained within both partial costs $J_{xp}^A$ and $J_{xp}^B$, thus, $J_{xp}^C \leq J_{xp}^A$ and $J_{xp}^C \leq J_{xp}^B$. Thus, the upper bounds holds true for any choice of orthonormal set of $\phi_x$ and $\phi_p$. Again, $J_{\mathrm{xp}}^C \geq 0$ because it is a sum of non-negative elements. □

**Theorem 4** (Bounds on Joint Projection Error for Greedy-Optimal Projections)**.** *Let $\phi_x$ and $\phi_p$ be chosen as the leading eigenvectors of $\widetilde{M}_x$ and $\widetilde{M}_p$ respectively, minimizing the partial costs $J_{\mathrm{xp}}^A$ and $J_{\mathrm{xp}}^B$. Then the joint cost $J_{\mathrm{xp}}$ satisfies the bounds*

$$\max\Big( \sum_{i=r_{\mathrm{x}}+1}^{n_{\mathrm{x}}} \lambda_i(\widetilde{M}_{\mathrm{x}}), \sum_{j=r_{\mathrm{p}}+1}^{n_{\mathrm{p}}} \lambda_j(\widetilde{M}_{\mathrm{p}}) \Big) \leq J_{\mathrm{xp}}(\phi_x, \phi_p) \leq \sum_{i=r_{\mathrm{x}}+1}^{n_{\mathrm{x}}} \lambda_i(\widetilde{M}_{\mathrm{x}}) + \sum_{j=r_{\mathrm{p}}+1}^{n_{\mathrm{p}}} \lambda_j(\widetilde{M}_{\mathrm{p}}).$$

*Proof.* By definition, the partial costs $J_{\mathrm{xp}}^A$ and $J_{\mathrm{xp}}^B$ are minimized by choosing $\phi_x$ and $\phi_p$ as the leading eigenvectors of $\widetilde{M}_x$ and $\widetilde{M}_p$, respectively. Thus, the minimal partial costs correspond to the sums of discarded eigenvalues:

$$J_{\mathrm{xp}}^A(\phi_x) = \sum_{i=r_{\mathrm{x}}+1}^{n_{\mathrm{x}}} \lambda_i(\widetilde{M}_x), \quad J_{\mathrm{xp}}^B(\phi_p) = \sum_{j=r_{\mathrm{p}}+1}^{n_{\mathrm{p}}} \lambda_j(\widetilde{M}_p).$$

From the decomposition of the joint cost $J_{\mathrm{xp}} = J_{\mathrm{xp}}^A + J_{\mathrm{xp}}^B - J_{\mathrm{xp}}^C$ and the bounds $0 \leq J_{\mathrm{xp}}^C \leq \min(J_{\mathrm{xp}}^A, J_{\mathrm{xp}}^B)$, it follows that

$$J_{\mathrm{xp}}^A + J_{\mathrm{xp}}^B - \min(J_{\mathrm{xp}}^A, J_{\mathrm{xp}}^B) \leq J_{\mathrm{xp}} \leq J_{\mathrm{xp}}^A + J_{\mathrm{xp}}^B,$$

which simplifies to

$$\max(J_{\mathrm{xp}}^A, J_{\mathrm{xp}}^B) \leq J_{\mathrm{xp}} \leq J_{\mathrm{xp}}^A + J_{\mathrm{xp}}^B.$$

Substituting the eigenvalue sums for $J_{\mathrm{xp}}^A$ and $J_{\mathrm{xp}}^B$ completes the proof. □

Finally, the projection matrices for the Petrov-Galerkin projection for LPV-SS proposed in Section 4 are defined to obtain the reduced model $\Sigma_r$ of the form (25), where:

$$V = \begin{pmatrix} \phi_1^{(x)} & \dots & \phi_{r_{\mathrm{x}}}^{(x)} \end{pmatrix} \in \mathbb{R}^{n_{\mathrm{x}} \times r_{\mathrm{x}}}, \quad Z = \begin{pmatrix} \phi_1^{(p)} & \dots & \phi_{r_{\mathrm{p}}}^{(p)} \end{pmatrix} \in \mathbb{R}^{n_{\mathrm{p}} \times r_{\mathrm{p}}}, \quad W = \begin{pmatrix} \phi_1^{(\Delta_x)} & \dots & \phi_{r_{\mathrm{x}}}^{(\Delta_x)} \end{pmatrix} \text{ or } W = V.$$

By analyzing the optimization of the coupled cost $J_{\mathrm{xp}}$, two approaches arise for choosing these projection vectors $\phi_x$ and $\phi_p$. One approach selects them as the dominant singular vectors of the tensor $\mathcal{M}$, while the other selects $\phi_x$ and $\phi_p$ as eigenvectors of the matrices $\widetilde{M}_x$ and $\widetilde{M}_p$, respectively, which represent appropriate marginalizations of $\mathcal{M}$ along the state and scheduling dimensions.

# 6 | SIMULATION STUDY

In this section, we analyse reduction performance, scalability, and computational load of the proposed joint reduction methods and we demonstrate their superior capabilities over existing methods. First, we introduce benchmark models and simulation data used for both reduction and validation together with performance metrics to evaluate the reduction methods. We use two different configurations of a nonlinear *mass-spring-damper* (MSD) system as benchmarks. The low complexity MSD benchmark with $n_{\mathrm{x}} = 10$ and $n_{\mathrm{p}} = 5$ is used to test the algorithm on a typical, manageable system size, where the relatively small number of scheduling variables allows comparison with a wide range of existing methods. To test scalability of the proposed approaches, a more complex MSD benchmark with $n_{\mathrm{x}} = 100$ and $n_{\mathrm{p}} = 99$ is also used.

For all examples, a laptop with Apple M3 chip and 16-GB RAM is used, running macOS and MATLAB 2024b. For comparison, we use the approaches implemented in LPVcore toolbox (v.0.10)[§], see[27] for the LPV embedding and independent state order or scheduling dimension reduction. For operations with tensors, we use the Tensor Toolbox for MATLAB by Sandia National Laboratories[28]. Lastly, for TSVD, a software developed in[21] is used, which is extended to the $N$-dimensional case.

## 6.1 | Benchmark Model

As a benchmark model, a chain of interconnected *mass-spring-damper* (MSD) blocks is used moving along a single dimension. All masses in the resulting MSD system are connected to an infinitely rigid wall with a spring and damper, but they are also connected with a spring and damper to the direct neighbours. The last mass in the chain is also effected by an external force denoted by $u$. Some of the springs are nonlinear in the interconnection while all other components are taken to be linear. The interconnection is depicted in Figure 1. The overall MSD system with $M$ blocks is described by the motion equations:

$$
\begin{aligned}
m_1\ddot{q}_1 &= -F_1 - F_{1,2} \\
m_i\ddot{q}_i &= -F_i - F_{i,i-1} - F_{i,i+1}, \quad i \in \mathbb{I}_2^{M-1} \\
m_M\ddot{q}_{\mathrm{M}} &= -F_{\mathrm{M}} - F_{\mathrm{M,M-1}} + u,
\end{aligned}
\tag{65}
$$

where $F_{i,j}$ is the force applied to the $i$-th mass by the adjacent masses, $F_i$ is the force applied to the $i$-th mass from the connection to the rigid wall, and $q_i$ and $\dot{q}_i$ represent the position and velocity of the $i$-th mass in meters and meters per second, respectively. The number of blocks that are connected with nonlinear springs is chosen as $M_{\mathrm{p}}$ and the resulting force expressions $F_{i,j}$ and $F_i$

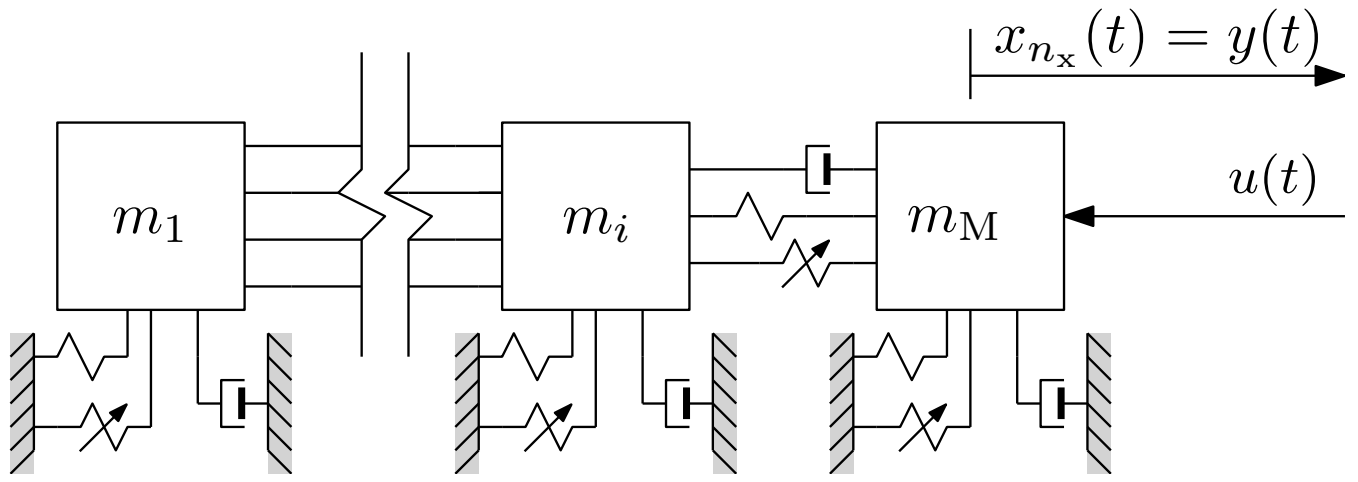

**FIGURE 1** Illustration of the interconnection of mass-spring-damper (MSD) systems with nonlinear elements.

[§] LPVcore toolbox available at https://lpvcore.net/

are as follows:

$$F_{i,j} = \begin{cases} k_\mathrm{a}(q_i - q_j) + k_\mathrm{b}(q_i - q_j)^3 + b(\dot{q}_i - \dot{q}_j), & \text{for the last } M_\mathrm{p} \text{ blocks, where nonlinear springs are present,} \\ k_\mathrm{a}(q_i - q_j) + b(\dot{q}_i - \dot{q}_j), & \text{for other blocks where the nonlinear springs are absent.} \end{cases}$$

Similarly, the force on the $i$-th mass $F_i$ is:

$$F_i = \begin{cases} k_\mathrm{a} q_i + k_\mathrm{b} q_i^3 + b\dot{q}_i, & \text{for the last } M_\mathrm{p} \text{ blocks,} \\ k_\mathrm{a} q_i + b\dot{q}_i, & \text{for other blocks.} \end{cases}$$

The parameters are chosen as $m_i = 0.1$ kg, for $i \in \mathbb{I}_1^M$, $k_\mathrm{a} = 0.5$ N/m, $k_\mathrm{b} = 0.5$ N/m, and $b = 1$ Ns/m. The external force $u$ is applied to the last mass in Newtons. Then, a general form of an NL-SS model of the system (65) where the output is the position of the last mass $y(t) = q_\mathrm{M}(t)$ is obtained.

We choose two different configurations of MSD systems. The first one, $\mathrm{MSD}_1$, contains $M = 5$ mass blocks, and $M_\mathrm{p} = 2$ blocks with nonlinear springs. The relatively small number of nonlinear springs allows us to test and compare the proposed methods with a TP model transformation approach. The second one, $\mathrm{MSD}_2$, contains $M = 50$ mass blocks, all of them connected with nonlinear springs ($M_\mathrm{p} = 50$). Both systems are converted into a global LPV-SS model using the so-called FTC approach[29] implemented in LPVcore, where the number of scheduling variables $n_\mathrm{p}$ corresponds to the number of nonlinear springs. The resulting LPV-SS models are as follows: MSD1 with $n_\mathrm{x} = 10$ and $n_\mathrm{p} = 5$, and MSD2 with $n_\mathrm{x} = 100$ and $n_\mathrm{p} = 99$.

## 6.2 | Simulation Data and Performance Metrics

We have generated trajectory data by executing time-domain simulations with the nonlinear benchmark models. For this purpose, we used the `ode45` solver with a sampling time of $T_d = 0.001$ s. Three different signals, $u_\mathrm{red}$, $u_\mathrm{val}$, and $u_\mathrm{extra}$, are used as inputs to the system. All three signals are a combination of step signals and multi-sine signals, but $u_\mathrm{extra}$ contains larger amplitudes, designed to excite the system into regions of the state-space not covered by $u_\mathrm{red}$ to test generalization capability. The initial conditions are set to $x_0 = 0$ for the each dataset and for both benchmark systems $\mathrm{MSD}_1$ and $\mathrm{MSD}_2$. The simulation trajectories are gathered to form two distinct datasets: $\mathcal{D}_\mathrm{red} = \{u_\mathrm{red}(t), x_\mathrm{red}(t), y_\mathrm{red}(t)\}_{t=0}^{N-1}$, $\mathcal{D}\mathrm{val} = \{u_\mathrm{val}(t), x_\mathrm{val}(t), y_\mathrm{val}(t)\}_{t=0}^{N-1}$, and $\mathcal{D}_\mathrm{extra} = \{u_\mathrm{extra}(t), x_\mathrm{extra}(t), y_\mathrm{extra}(t)\}_{t=0}^{N-1}$. The dataset $\mathcal{D}_\mathrm{red}$ is used for the data-based reduction methods, while $\mathcal{D}_\mathrm{val}$ is reserved for validation purposes. The input trajectories are shown in Fig.2.a, and Fig.2.b shows the simulated output trajectories of the full-order model $\mathrm{MSD}_1$. A visualization of how $\mathcal{D}_\mathrm{extra}$ explores a larger region of the state space compared to $\mathcal{D}_\mathrm{red}$, and $\mathcal{D}_\mathrm{val}$ is shown in Fig. 2.c.

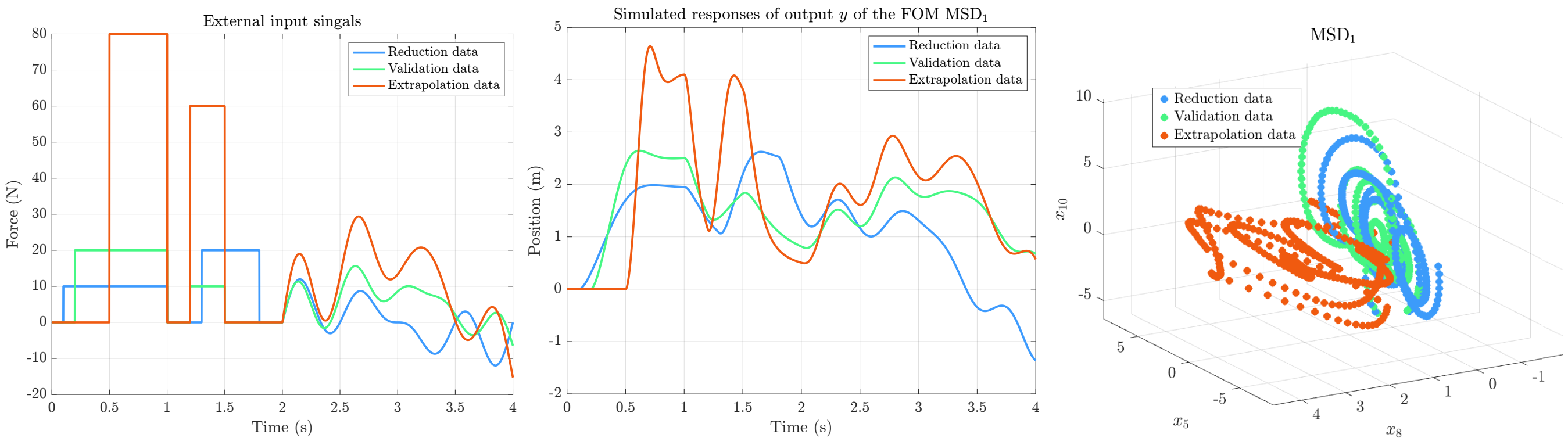

(a) (b) (c)

**FIGURE 2** Datasets obtained for $\mathrm{MSD}_1$: (a) reduction, validation, and extrapolation input trajectories; (b) obtained output trajectories; (c) state region represented by the data.

The first performance metric that we consider is the *normalized root-mean-square error* (NRMSE) of the simulated model response w.r.t. the output of the original system , given as

$$\text{NRMSE} = \frac{\|y - \hat{y}\|_{\ell_2}}{\|y - y_{\text{mean}}\|_{\ell_2}} \cdot 100\%, \tag{66}$$

where $\{y(tT_\text{s})\}_{t=0}^{N-1}$, $\{\hat{y}(tT_\text{s})\}_{t=0}^{N-1}$ are the respective output signals of the original and reduced model, sampled at time instances $tT_\text{s}$ and $y_\text{mean}$ is the mean of the sampled sequence $y$. The second performance metric we consider is the *computation time* (CPU) required to execute a reduction method on a given benchmark. CPU time measurements in this study were performed using `tic` and `toc` functions built-in MATLAB. Additionally, for the POD method, we evaluate the $\ell_2$ norms of the state and scheduling variable errors, introduced in Subsection 5.2 and denoted as $J_\text{x}$, $J_\text{p}$, and $J_\text{xp}$, calculated both at the signal level and with dynamics included, to provide a comprehensive assessment of the reduction performance.

## 6.3 | Tensor LPV Moment Matching Analysis

First we give an in-depth analysis of the tensor-based LPV moment matching method on the $\text{MSD}_1$ benchmark under various scenarios. In the first scenario, the horizon level is fixed at $n = 2$ and the reachability and observability tensors are decomposed using the TSVD tensor decomposition. The purpose of this scenario is to compare the differences between the 3 modes of the method. Fig. 3.a presents the output responses of the resulting reduced models, simulated with an input signal from $\mathcal{D}_\text{val}$, alongside the absolute error trajectories $e_\text{val}(t) = \left|y_\text{val}(t) - \hat{y}_\text{val}(t)\right|$. It is observed that, for $\text{MSD}_1$, the observability mode yields a marginally lower NRMSE; the Hankel mode better captures step responses but struggles with the multi-sine signals, while the reachability-based reduction introduces the largest error because the dynamics captured with the reachability mode with horizon $n = 2$ are insufficient to accurately describe the system. The characteristics of the resulting reduced models are summarized in Table 2, where it is shown that the proposed TMM effectively reduces the number of states from $n_\text{x} = 10$ to $r_\text{x} = 3$ and the number of scheduling variables from $n_\text{p} = 5$ to $r_\text{p} = 2$.

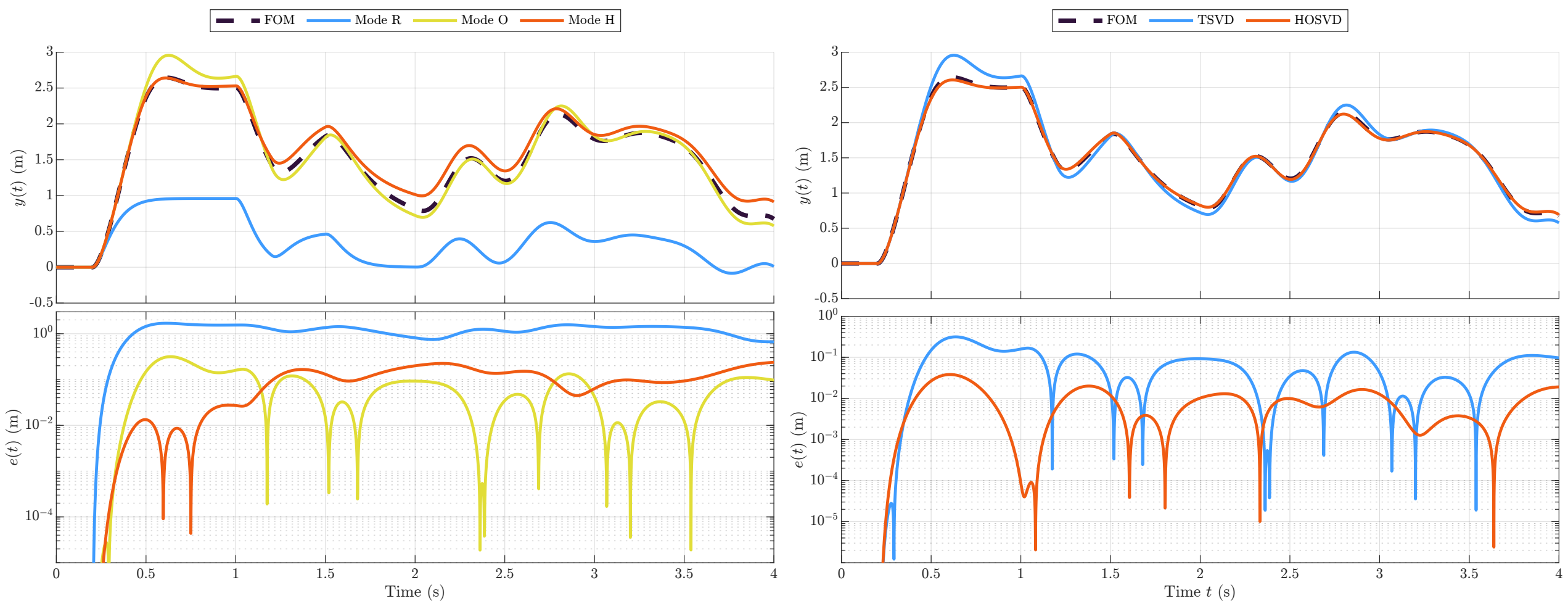

(a) (b)

**FIGURE 3** Accuracy of the obtained reduced models with TMM for $\text{MSD}_1$ in terms of simulated output response on the validation data set: (a) comparison of reachability, observability, and Hankel modes with TSVD decomposition and $n = 2$; (b) comparison of TSVD and HOSVD decomposition in the observability mode with $n = 2$.

In the second scenario, we analyze the influence of choosing a different type of tensor decomposition during the model reduction process. For this scenario, we use the observability-based approach as it best captures the dynamics of $\text{MSD}_1$ and a fixed horizon of $n = 2$. Fig. 3.b depicts the output responses and the associated error trajectories of the resulting reduced models. Quantitive evaluation of the reduction performance is given in Table 2. It can be observed that choosing the HOSVD type of decomposition results in a higher number of singular vectors per horizon. Consequently, the reduced model obtained using HOSVD retains more states and scheduling variables compared to the model obtained using the TSVD type of decomposition. As a result, HOSVD achieves a lower NRMSE by preserving more system information. However, TSVD offers better compression by reducing system complexity further. Thus, the choice between HOSVD and TSVD depends on the trade-off between accuracy and compression.

**TABLE 2** Reduction performance of the TMM method with different modes and decompositions and horizon $n = 2$ on $\text{MSD}_1$.

| Model | $n_\text{x}$ | $n_\text{p}$ | NRMSE (%) | CPU (s) |
|---|---|---|---|---|
| Original model | 10 | 5 | - | - |
| Reachability mode (with TSVD) | 3 | 2 | 183.14 | 0.05 |
| Reachability mode (with HOSVD) | 5 | 4 | 0.48 | 0.05 |
| Observability mode (with TSVD) | 3 | 2 | 15.28 | 0.19 |
| Observability mode (with HOSVD) | 4 | 4 | 1.84 | 0.02 |
| Hankel matrix mode (with TSVD) | 3 | 3 | 18.23 | 0.12 |
| Hankel matrix mode (with HOSVD) | *Rank condition not satisfied.* | | | |

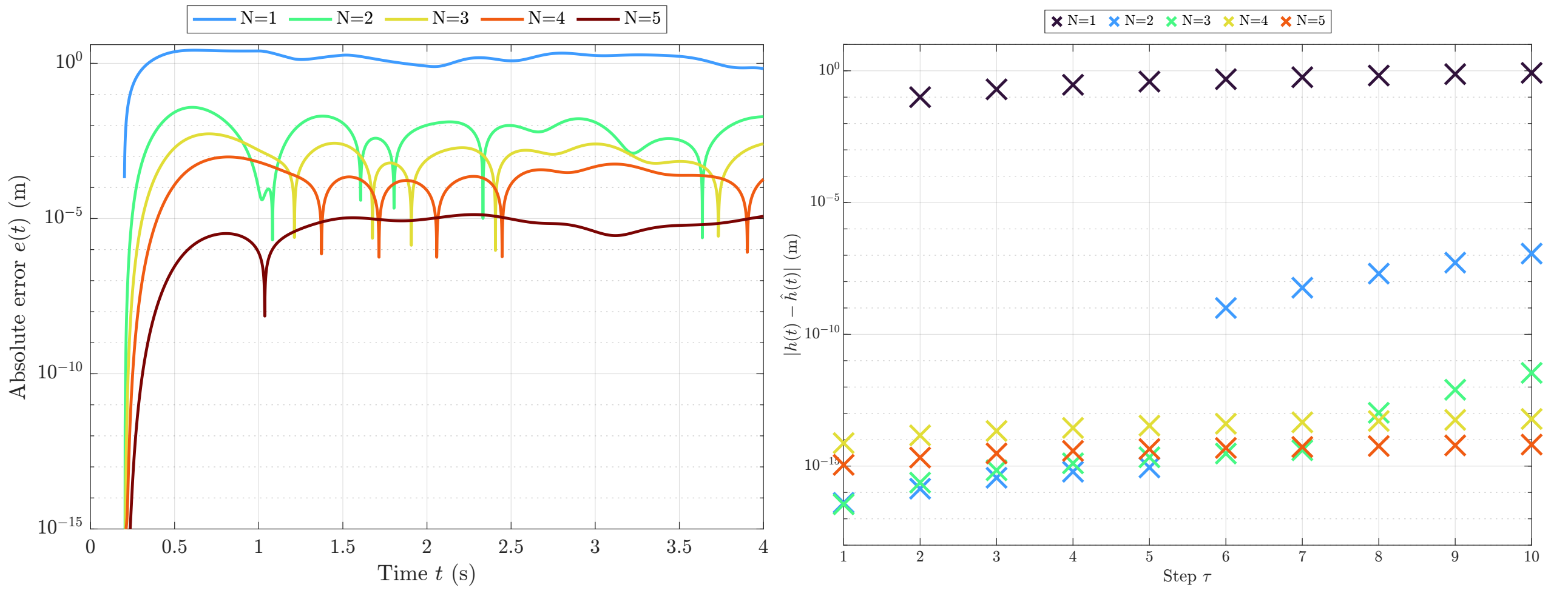

(a) (b)

**FIGURE 4** Accuracy of the obtained reduced models with TMM for $\text{MSD}_1$ in terms of simulated output response on the validation data set: (a) reachability mode with TSVD under various horizon levels $n$; (b) Impulse response error $\left|h(\tau) - \hat{h}(\tau)\right|$ of the reduced models at various horizon levels $n$.

The third scenario examines how increasing the horizon level $n$ affects model reduction. Table 3 summarizes key performance metrics, including reachability tensor properties and the impact of using TSVD vs. HOSVD. Since TSVD approximates solutions as sum of rank-1 tensors, achieving an exact representation is often infeasible. To quantify this, we report the Frobenius norm error $\|\mathcal{R}_\text{n} - \hat{\mathcal{R}}_\text{n}\|_F$, along with singular values $\sigma(\mathcal{R}_\text{n})$, CPU time, reduced model dimensions $(r_\text{x}, r_\text{p})$, and NRMSE of output reponses on the validation data. Since the original model has full-rank reachability and observability matrices for $n = n_\text{x}$, the reduction process balances accuracy vs. complexity. Even for $n = 2$, using HOSVD with the reachability mode, we reduced the original model from $n_\text{x} = 10$ to $r_\text{x} = 5$ and $n_\text{p} = 5$ to $r_\text{p} = 4$, achieving an NRMSE of only 0.48%. For HOSVD, the number of additional singular vectors in state and scheduling dimensions grows with $N$, leading to faster model order expansion. The results reaffirm results of the second scenario: HOSVD yields more accurate models, while TSVD achieves stronger compression. Lastly in Fig. 4.a, a gradual decrease in the absolute error can be observed as the horizon level $n$ increases. The impulse responses for both the original and reduced models are obtained as output trajectories using a Dirac impulse input, $u(t) = \delta(t)$. The impulse response error signal, depicted in Fig. 4.b, shows that models reduced with larger $n$ values exhibit nearly perfect matching of higher number of impulse response coefficients.

**TABLE 3** Analysis of reachability-based TMM applied to the $\text{MSD}_1$ model.

| | TSVD | | | | | | HOSVD | | | | |
|---|---|---|---|---|---|---|---|---|---|---|---|
| **Horizon** | $\|\mathcal{R}_n - \hat{\mathcal{R}}_n\|_F$ | $\sigma(\mathcal{R}_n)$ | CPU (s) | $r_x$ | $r_p$ | NRMSE (%) | Core dimension | CPU (s) | $r_x$ | $r_p$ | NRMSE (%) |
| $n = 0$ | 0 | 0.0100 | 0.04 | 1 | 0 | 246.89 | 1 | 0.016 | 1 | 0 | 246.89 |
| $n = 1$ | 0 | 0.0098 | 0.05 | 2 | 0 | 246.89 | 1 | 0.019 | 2 | 0 | 246.89 |
| $n = 2$ | $1 \times 10^{-7}$ | 0.0096 | 0.06 | 3 | 2 | 183.14 | $(3, 3)$ | 0.027 | 5 | 4 | **0.48** |
| $n = 3$ | $1 \times 10^{-7}$ | 0.0094 | 0.06 | 4 | 3 | 136.47 | $(4, 4, 3)$ | 0.032 | 7 | 5 | 0.13 |
| $n = 4$ | $2 \times 10^{-7}$ | 0.0092 | 79.89 | 5 | 4 | 48.54 | $(5, 4, 4, 3)$ | 0.033 | 8 | 6 | 0.01 |
| $n = 5$ | $3 \times 10^{-7}$ | 0.0091 | 105.11 | 6 | 5 | 1.33 | $(5, 4, 4, 4, 3)$ | 0.76 | 9 | 6 | 0 |
| $n = 6$ | $5 \times 10^{-7}$ | 0.0089 | 106.32 | 6 | 6 | 1.31 | $(5, 5, 4, 5, 4, 3)$ | 0.24 | 10 | 6 | 0 |

## 6.4 | Analysis of POD Approach to LPV Model Reduction

In this section, we present the results of applying the POD approach to model $\text{MSD}_1$. Based on the reduction dataset $\mathcal{D}_{\text{red}}$, we constructed the matrices $M_x$, $M_p$, $\widetilde{M}_x$, $\widetilde{M}_p$, and the tensor $\mathcal{M}$ as described in Section 5.2.2.

We distinguish four approaches to obtain the POD reduction. Among these approaches, only the first method treats the projections independently, while the remaining three preserve the coupling by jointly computing the projections. Specifically:

1. **Matrix case**: Employs singular vectors of $M_x$ for state projection and singular vectors of $M_p$ for scheduling projection.
2. **Weighted matrix case**: Uses eigenvectors of $\widetilde{M}_x$ and $\widetilde{M}_p$ to find $V$ and $Z$, which depend on both $x$ and $p$.
3. **Tensor case (TSVD)**: Uses singular vectors obtained from tensor $\mathcal{M}$ via TSVD.
4. **Tensor case (HOSVD)**: Similar to the previous method but uses HOSVD.

First, we aim to determine reasonable choices for $r_x$ and $r_p$. Figure 5 shows the values of the optimal normalized cost functions $J_x$, $J_p$, $J_{xp}^A$, and $J_{xp}^B$, all calculated using the reduction dataset, as defined in Section 5.2. As a reminder, $J_x$ and $J_p$ correspond to the matrix case, while $J_{xp}^A$ and $J_{xp}^B$ correspond to the weighted matrix case. These values are obtained based on projections only, meaning that $\hat{x}(t) = VV^\top x(t)$ and $\hat{p}(t) = ZZ^\top p(t)$. From the plots, it can be inferred that choosing $r_x = 3$ and $r_p = 2$ is reasonable, since the error drops below 1% for larger values.

The next objective is to analyze the impact of independently versus jointly computing the projection matrices $V$, $W$, and $Z$. To this end, we applied all four variations of the reduction method using the previously selected values $r_x = 3$ and $r_p = 2$, and evaluated the state cost $J_x$, scheduling cost $J_p$, and joint cost $J_{xp}$ across the reduction, validation, and extrapolation datasets. The corresponding values, obtained from the full simulation of the reduced model (including dynamics and projections, i.e., $\hat{x}(t) = Vx_r(t)$ and $\hat{p}(t) = Zp_r(t)$), are presented in Table 4. Additionally, Fig. 6 depicts the output responses and the associated error trajectories of the resulting reduced models on $\mathcal{D}_{\text{val}}$ and $\mathcal{D}_{\text{extra}}$.

Several trends emerge from the results:

- **Joint projections outperform independent ones.** In all datasets, methods that preserve the coupling between the state and scheduling subspaces yield lower costs, confirming the benefit of joint reduction strategies.

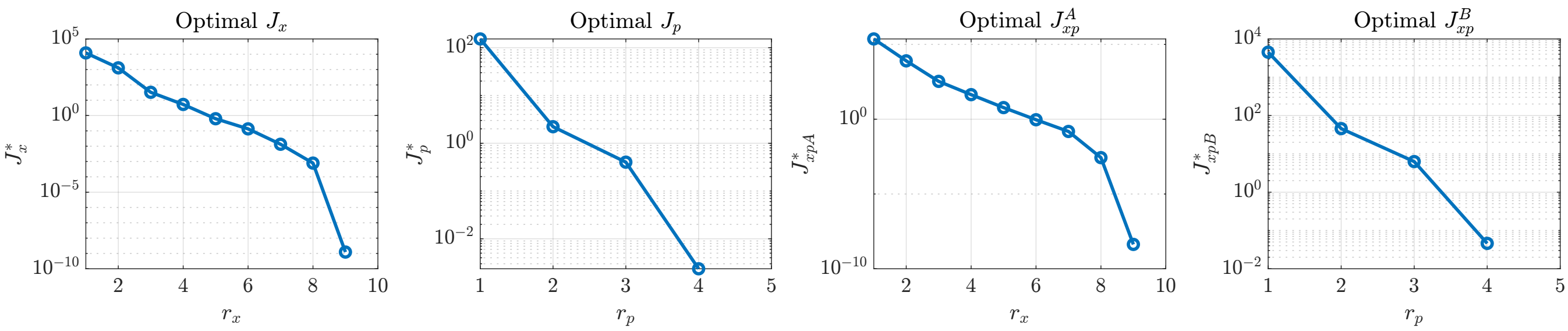

**FIGURE 5** Normalized optimal values of the cost functions for different $r_x$ and $r_p$, showing how the choice of reduction dimensions influences the objectives $J_x$, $J_p$, $J_{xp}^A$, and $J_{xp}^B$.

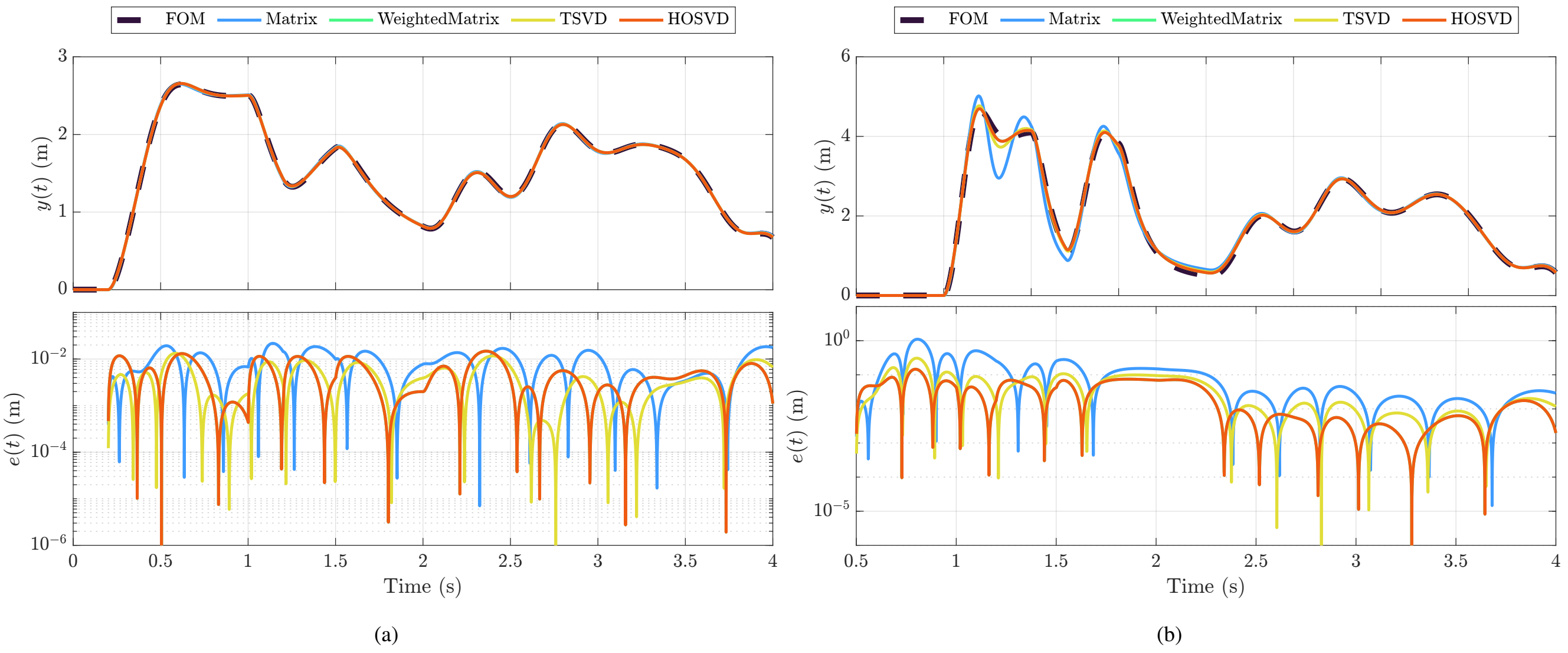

**FIGURE 6** Comparison of reduced models for $\mathrm{MSD}_1$ with $r_x = 3$ and $r_p = 2$: (a) Output and absolute error plots on the validation dataset; (b) Output and absolute error plots on the extrapolation dataset.

- **HOSVD and Weighted Matrix yield identical subspaces.** Despite originating from different formulations, both approaches consistently result in the same projection spaces and identical cost values, as confirmed by direct comparison of the resulting bases.
- **HOSVD/Weighted Matrix perform best overall.** They achieve the lowest costs across all metrics in both reduction and extrapolation datasets, while in the validation dataset, TSVD slightly outperforms them.
- **Matrix method fails under extrapolation.** While the Matrix method performs reasonably under reduction and validation conditions, it breaks down significantly in extrapolation, indicating its poor generalization to unseen data.

As explained in Section 5.2.2, weighted matrix case is the approach that emerged naturally as a sub-optimal solution, while minimising the parts A and B of the joint cost $J_{\mathrm{xp}}$. While the optimal solution is not explicitly found, the SVD of tensor $\mathcal{M}$ is one approach to find it. A surprising observation is that HOSVD provides equivalent results to the weighted matrix case, indicating an intrinsic connection between the methods. Consequently, it means that there is a computationally less complex way of performing this tensor-based POD approach of model reduction, using the matrices $\widetilde{M}_x$ and $\widetilde{M}_{\mathrm{p}}$.

## 6.5 | Scalability Analysis on $\mathrm{MSD}_2$

In this subsection, we evaluate the scalability of the proposed tensor-based methods by applying them to the $\mathrm{MSD}_2$ benchmark, which features a significantly larger number of states ($n_{\mathrm{x}} = 100$) and scheduling variables ($n_{\mathrm{p}} = 99$).

The results obtained with the TMM method using reachability-based reduction with $n = 2$ and the POD method using Galerkin projection are summarized in Table 5. The results are presented both with TSVD and HOSVD decompositions. The last column shows the total number of model parameters, which is the sum of the number of parameters from the tensors $\mathcal{A}$, $\mathcal{B}$, $\mathcal{C}$, and $\mathcal{D}$, providing another measure of reduction effectiveness.

**TABLE 4** Cost function and NRMSE values for Reduction, Validation, and Extrapolation datasets

| Method | Reduction | | | | Validation | | | | Extrapolation | | | |
|---|---|---|---|---|---|---|---|---|---|---|---|---|
| | $J_{\mathrm{x}}$ | $J_{\mathrm{p}}$ | $J_{\mathrm{xp}}$ | NRMSE | $J_{\mathrm{x}}$ | $J_{\mathrm{p}}$ | $J_{\mathrm{xp}}$ | NRMSE | $J_{\mathrm{x}}$ | $J_{\mathrm{p}}$ | $J_{\mathrm{xp}}$ | NRMSE |
| Matrix | 118.43 | 47.81 | 4029 | 1.84 | 85.97 | 31.41 | 3299 | 1.47 | $1.88 \cdot 10^5$ | $4.01 \cdot 10^4$ | $1.13 \cdot 10^8$ | 16.07 |
| WeightedMatrix | **87.24** | **28.98** | **1284** | **1.12** | **64.28** | **11.01** | 1498 | 0.91 | $\mathbf{2.05 \cdot 10^4}$ | $\mathbf{5.29 \cdot 10^3}$ | $\mathbf{8.84 \cdot 10^6}$ | **3.14** |
| TSVD | 93.27 | 30.71 | 1680 | 1.23 | 64.73 | 11.14 | **1366** | **0.76** | $4.20 \cdot 10^4$ | $9.96 \cdot 10^3$ | $1.88 \cdot 10^7$ | 5.10 |
| HOSVD | **87.24** | **28.98** | **1284** | **1.12** | **64.28** | **11.01** | 1498 | 0.91 | $\mathbf{2.05 \cdot 10^4}$ | $\mathbf{5.29 \cdot 10^3}$ | $\mathbf{8.84 \cdot 10^6}$ | **3.14** |

**TABLE 5** Scalability analysis of tensor-based reduction methods applied to $MSD_2$

| Model | $n_x$ | $n_p$ | NRMSE (%) | CPU (s) | # Parameters |
|---|---|---|---|---|---|
| Original model | 100 | 99 | - | - | $1.02 \times 10^6$ |
| TMM with TSVD | 3 | 2 | 15.29 | 10.40 | 32 |
| TMM with HOSVD | 4 | 4 | 1.84 | 1.22 | 100 |
| POD with TSVD | 3 | 2 | 0.76 | $1.54 \cdot 10^3$ | 48 |
| POD with HOSVD | 3 | 2 | 0.89 | 4.25 | 48 |

**TABLE 6** Dimensions of reachability and observability tensors of $MSD_2$

| Horizon | Reachability tensor | Observability tensor |
|---|---|---|
| $n = 0$ | $(100, 100, 1)$ | $(1, 100, 100)$ |
| $n = 1$ | $(100, 100, 100, 1)$ | $(1, 100, 100, 100)$ |
| $n = 2$ | $(100, 100, 100, 100, 1)$ | $(1, 100, 100, 100, 100)$ |
| $n = 3$ | $(100, 100, 100, 100, 100, 1)$ | $(1, 100, 100, 100, 100, 100)$ |

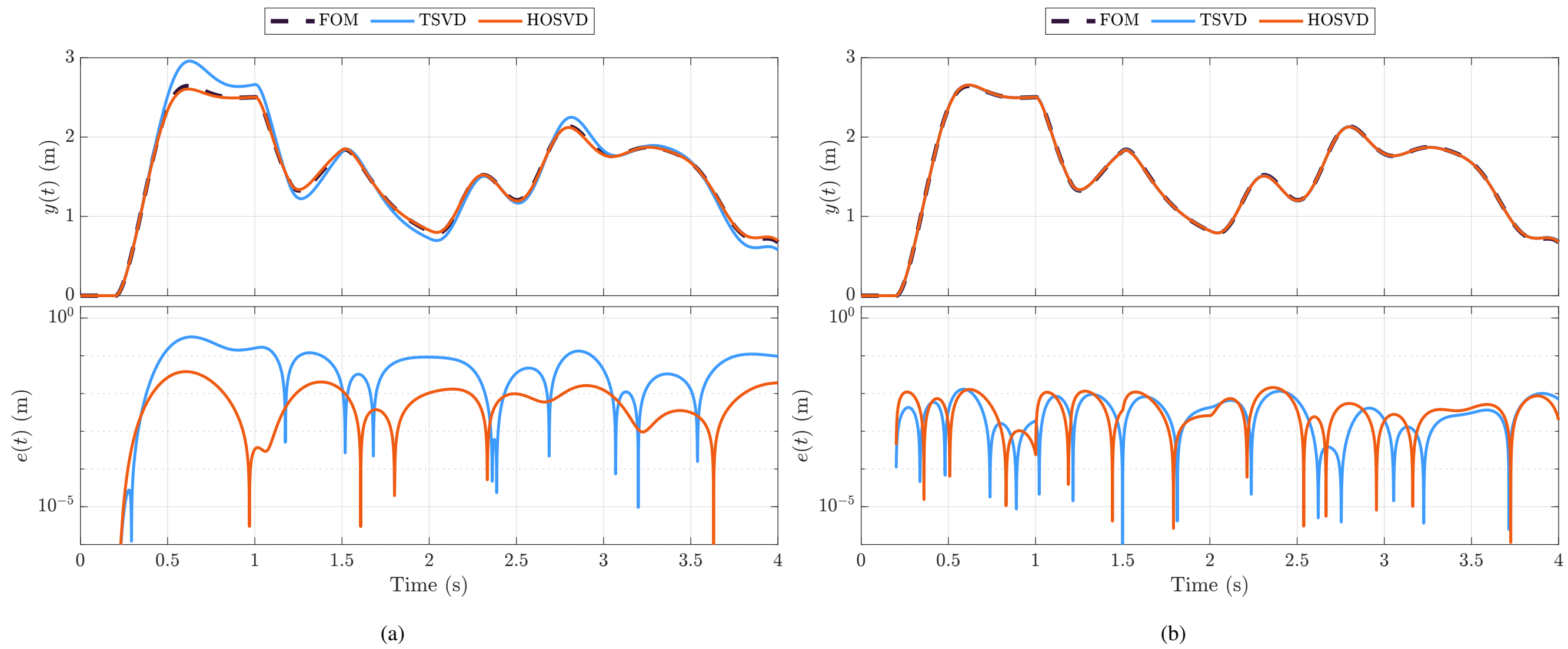

**FIGURE 7** Accuracy of the obtained reduced models for $MSD_2$ in terms of simulated output response on the validation data set: (a) reachability-based TMM; (b) POD with Galerkin projection.

Both methods achieve significant compression, but with different characteristics. TMM, using TSVD and HOSVD, reduces the original model, containing over $1.02 \times 10^6$ parameters, to just 32 and 100 parameters, respectively. This drastic reduction is due to the dominance of a few key states and scheduling variables in the overal systme behavior. POD, while operating on data-driven principles, achieves even lower scheduling dimensions with a low NRMSE less than $1\%$. However, both approaches face computational scalability limits. Table 6 highlights how for TMM tensor sizes grow with increasing $n$, making large-horizon reductions computationally expensive due to the rapid increase in reachability and observability tensors and decomposition complexity.

While both TMM and POD can handle high-dimensional systems and preserve essential dynamics effectively, their scalability is primarily influenced by tensor size and sparsity, which affect memory usage and decomposition time. Nevertheless, compared to traditional methods, they remain among the few feasible approaches for reducing large-scale LPV models.

## 6.6 | Comparative Analysis of LPV Model Reduction Methods

In this subsection, we compare our proposed methods with other LPV model reduction techniques applied on $MSD_1$. The purpose of this comparison is to assess the effectiveness of the proposed tensor-based methods against established techniques that operate independently on state and scheduling reduction. Each of these methods is tested on $MSD_1$ and we compare their performance based on reduction effectiveness, computational cost, and the preservation of key characteristics of the system behavior. We evaluate the following methods:

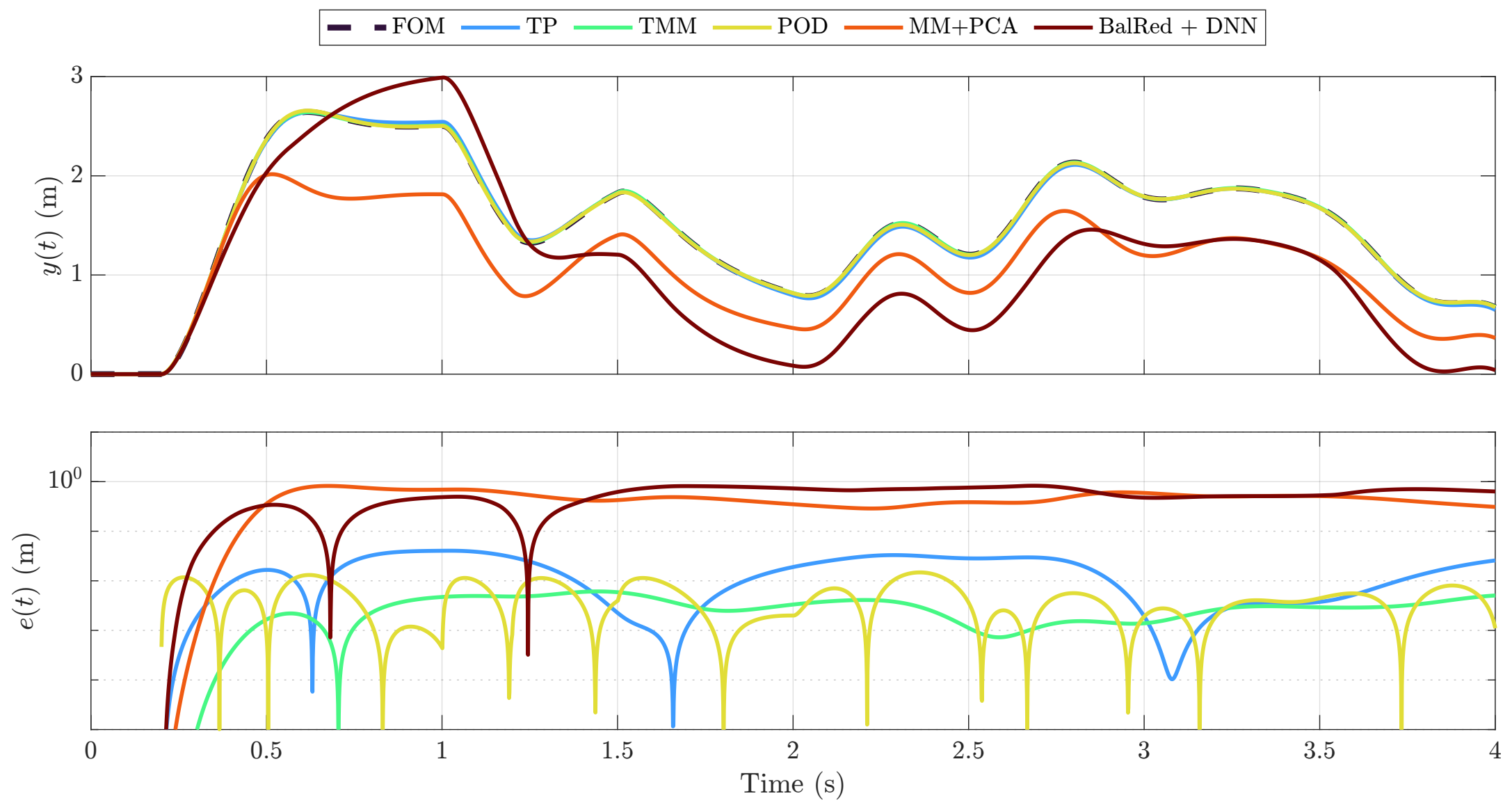

**FIGURE 8** Comparison of the obtained reduced models by state-of-the-art methods for $\mathrm{MSD}_1$ in terms of simulated output response on the validation dataset. The full-order model is a black dashed line. Reduced models: tensor-product model transformation (blue), joint tensor moment matching (green), joint tensor POD with HOSVD (yellow), independent moment matching and PCA (red), and independent balanced reduction with DNN-based scheduling reduction (brown).

1. *Tensor-product model transformation* (TP method)[18,30], which employs HOSVD to reduce the number of vertices in the convex polytopic representation of an LPV-SS model. The method reduces the scheduling variables by reducing the number of vertices, but it does not support state reduction. For the $\mathrm{MSD}_1$ model, the full-order TP model initially has $2^5$ vertices, which is reduced to $2^2$, lowering the scheduling dimension to $r_\mathrm{p} = 2$.
2. *Tensor LPV moment matching* (TMM) (proposed method in Sec. 5.1), used in a reachability mode with HOSVD decomposition and horizon $n = 2$. It reduces the state dimension from $n_\mathrm{x} = 10$ to $r_\mathrm{x} = 5$ and the scheduling dimension from $n_\mathrm{p} = 5$ to $r_\mathrm{p} = 4$.
3. *Tensor proper orthogonal decomposition* (POD) (proposed method in Sec. 5.2), used with Galerkin projection with HOSVD type of decomposition, and the selected values of state order and scheduling dimension are $r_\mathrm{x} = 3$, and $r_\mathrm{p} = 2$.
4. State reduction with *moment matching*[7] combined with *principal-component-analysis* (PCA)-based scheduling reduction[10]. In this approach, the state dimension is first reduced using matrix-based moment matching, followed by the reduction of the scheduling dimension using PCA. For $\mathrm{MSD}_1$, moment matching reduces the state dimension to $r_\mathrm{x} = 4$, and PCA reduces the scheduling dimension to $r_\mathrm{p} = 2$.
5. *Balanced reduction*[5] combined with *deep-neural-network*-based scheduling reduction[13] (Bal Red + DNN), corresponding to a method combining balanced reduction for state-order reduction with a DNN learning method for scheduling reduction. The state dimension are reduced from $n_\mathrm{x} = 10$ to $r_\mathrm{x} = 5$, followed by a scheduling reduction from $n_\mathrm{p} = 5$ to $r_\mathrm{p} = 2$.

The achieved performance of these model reduction methods is depicted in Fig. 8, which shows the output and error trajectories for $\mathrm{MSD}_1$ using $\mathcal{D}_\mathrm{val}$. We can observe that the joint reduction methods, such as TMM, outperform both sequential state-and-scheduling reduction approaches as well as the TP model transformation, which only reduces the scheduling dimension while leaving the state dimension unchanged. The tensor-based methods achieve the best balance between model size reduction and preservation of system dynamics, offering the most effective reduction while maintaining high accuracy.

## 7 | CONCLUSION

This paper advances the state-of-the-art in LPV model reduction by introducing the first systematic joint reduction approach for both state order and scheduling dimensions. The projection framework developed here, based on Petrov-Galerkin projections, provides a systematic method for reducing both the state-space $\mathbb{X}$ and scheduling space $\mathbb{P}$, constrained to linear transformations. A key contribution is the development of a tensor-based LPV moment matching method, an extension of the matrix-based moment matching method, which offers a powerful tool for jointly reducing non-minimal LPV models and preserving reachable and/or observable subspaces. We also show that there are specific reachability and observability subspaces that are connected to the variation scheduling variables, beyond the well-known reachability and observability subspaces describing connected to the state. A second contribution is the generalization of proper orthogonal decomposition for tensor-based reduction of LPV models, giving a combined data-driven sate and scheduling reduction method. For the developed methods, comparison of tensor decomposition methods shows that HOSVD is a more efficient and flexible approach, providing faster convergence and more consistent results than TSVD. While TSVD is optimal for diagonalizable tensors, it faces scalability challenges and higher computational demands, making HOSVD often a better choice in practice. Our results pave the way for the development of future generation of efficient reduced-order modelling approaches based on joint reduction, potentially enabling even nonlinear transformations of the state and scheduling variables, for instance, through kernelization and neural-networks-based learning methods.

## ACKNOWLEDGMENTS

This research was supported by The MathWorks Inc., the European Space Agency (ESA) through Initial Support for Innovation (EISI) in the Discovery Programme (contract: 4000145530), and the European Union within the framework of the National Laboratory for Autonomous Systems (RRF-2.3.1-21-2022-00002). Opinions, findings, conclusions or recommendations expressed in this paper are those of the authors and do not necessarily reflect the views of the The MathWorks Inc., European Space Agency, or the European Union.

# 8 | APPENDIX

## 8.1 | Proof of the optimal state approximation using POD

By manipulating the expression we get

$$
\begin{aligned}
J_x(\varphi_1^{(x)}, \dots, \varphi_{r_x}^{(x)}) &= \sum_{j=0}^{N-1} \|x(j) - \hat{x}(j)\|^2 = \\
&= \sum_{j=0}^{N-1} \| \sum_{k=1}^{n_x} \langle x(j), \varphi_k^{(x)} \rangle \varphi_k^{(x)} - \sum_{k=1}^{r_x} \langle x(j), \varphi_k^{(x)} \rangle \varphi_k^{(x)} \|^2 = \\
&= \sum_{j=0}^{N-1} \| \sum_{k=r_x+1}^{n_x} \langle x(j), \varphi_k^{(x)} \rangle \varphi_k^{(x)} \|^2 = \\
&= \sum_{j=0}^{N-1} \Big\langle \sum_{k=r_x+1}^{n_x} \langle x(j), \varphi_k^{(x)} \rangle \varphi_k^{(x)}, \sum_{l=r_x+1}^{n_x} \langle x(j), \varphi_l^{(x)} \rangle \varphi_l^{(x)} \Big\rangle = \\
&= \sum_{j=0}^{N-1} \sum_{k=r_x+1}^{n_x} \sum_{l=r_x+1}^{n_x} \langle x(j), \varphi_k^{(x)} \rangle \langle x(j), \varphi_l^{(x)} \rangle \langle \varphi_k^{(x)}, \varphi_l^{(x)} \rangle = \\
&= \sum_{j=0}^{N-1} \sum_{k=r_x+1}^{n_x} \langle x(j), \varphi_k^{(x)} \rangle^2 = \sum_{k=r_x+1}^{n_x} \sum_{j=0}^{N-1} \langle x(j), \varphi_k^{(x)} \rangle^2 = \\
&= \sum_{k=r_x+1}^{n_x} \sum_{j=0}^{N-1} (\varphi_k^{(x)})^\top x(j) x^\top(j) (\varphi_k^{(x)}) = \sum_{k=r_x+1}^{n_x} (\varphi_k^{(x)})^\top M_x M_x^\top (\varphi_k^{(x)})
\end{aligned}
$$

Then, the optimal approximation error of state projection is:

$$
J_x^*(\varphi_1^{(x)}, \dots, \varphi_{r_x}^{(x)}) = \sum_{k=r_x+1}^{n_x} (\varphi_k^{(x)})^\top M_x M_x^\top (\varphi_k^{(x)}) = \sum_{k=r_x+1}^{n_x} (\sigma_k^{(x)})^2 (\varphi_k^{(x)})^\top \varphi_k^{(x)} = \sum_{k=r_x+1}^{R_x} (\sigma_k^{(x)})^2
$$